%% file: planet_on_comets.tex
\documentclass[fleqn,usenatbib]{mnras}

\usepackage{newtxtext,newtxmath}
\usepackage[T1]{fontenc}
\usepackage{ae,aecompl}

\DeclareRobustCommand{\VAN}[3]{#2}
\let\VANthebibliography\thebibliography
\def\thebibliography{\DeclareRobustCommand{\VAN}[3]{##3}\VANthebibliography}

\usepackage{graphicx}	%
\usepackage{amsmath}	%
\usepackage{enumitem} %
\usepackage{svg}
\usepackage{float}

\usepackage{orcidlink}

\input{sections/commands}
\title[Planetary debris in star clusters]{Influence of planets on debris disks in star clusters I: the 50 AU Jupiter} %
\author[Wu et al.]{
Kai Wu\orcidlink{0000-0003-0349-0079}$^{1,2}$,
M.B.N. Kouwenhoven\orcidlink{0000-0002-1805-0570}$^{1}$\thanks{E-mail: t.kouwenhoven@xjtlu.edu.cn}, 
Rainer Spurzem\orcidlink{0000-0003-2264-7203}$^{3,4,5}$, 
Xiaoying Pang\orcidlink{0000-0003-3389-2263}$^{1}$
\\
$^{1}$Department of Physics, School of Mathematics and Physics, Xi'an Jiaotong-Liverpool University, 111 Ren'ai Rd., Industrial Park District,\\ Suzhou, Jiangsu 215123, China\\
$^{2}$Department of Mathematical Sciences, University of Liverpool, Liverpool L69 3BX, UK\\
$^{3}$Astronomisches Rechen-Institut, Zentrum f\"{u}r Astronomie der Universit\"{a}t  Heidelberg, M\"onchhofstr. 12-14, D-69120 Heidelberg, Germany\\
$^{4}$National Astronomical Observatories and Key Laboratory of Computational Astrophysics, Chinese Academy of Sciences, 20A Datun Rd.,\\ Chaoyang District, 100101, Beijing, China\\
$^{5}$Kavli Institute for Astronomy and Astrophysics, Peking University, Yiheyuan Lu 5, Haidian Qu, 100871, Beijing, China\\
}

\date{Accepted --; Received --; in original form --}

\pubyear{2023}

\begin{document}
\label{firstpage}
\pagerange{\pageref{firstpage}--\pageref{lastpage}}
\maketitle

\begin{abstract}
Although debris disks may be common in exoplanet systems, only a few systems are known in which debris disks and planets coexist. Planets and the surrounding stellar population can have a significant impact on debris disk evolution. Here we study the dynamical evolution of debris structures around stars embedded in star clusters, aiming to determine how the presence of a planet affects the evolution of such structures. We combine \texttt{NBODY6++GPU} and \texttt{REBOUND} to carry out \textit{N}-body simulations of planetary systems in star clusters ($N=8\,000$; $R_\mathrm{h}=0.78$~pc) for a period of 100~Myr, in which 100 solar-type stars are assigned 200 test particles. Simulations are carried out with and without a Jupiter-mass planet at 50~au. We find that the planet destabilizes test particles and speeds up their evolution. The planet expels most particles in nearby and resonant orbits. Remaining test particles tend to retain small inclinations when the planet is present, and fewer test particles obtain retrograde orbits. Most escaping test particles with speeds smaller than the star cluster's escape speed originate from cold regions of the planetary system or from regions near the planet. We identify three regions within planetary systems in star clusters: (i) the private region of the planet, where few debris particles remain ($40-60$~au), (ii) the reach of the planet, in which particles are affected by the planet ($0-400$~au), and (iii) the territory of the planetary system, most particles outside which will eventually escape ($0-700$~au). 
\end{abstract}

\begin{keywords}
planetary systems - stars: solar-type - stars: statistics - methods: numerical - planets and satellites: dynamical evolution and stability - galaxies: star clusters: general
\end{keywords}

\raggedbottom %
\input{sections/intro}

\input{sections/method}

\input{sections/result}

\input{sections/conclusion}

\section*{Acknowledgements}

We are grateful to the anonymous referee for providing comments and suggestions that helped to improve this paper.
We are grateful to Tai-Jun Chen and Martin Gorbahn for the discussions that helped to improve the paper.
KW thanks Mingze Sun, Francesco M. Flammini Dotti, and Xiuming Xu for their helpful discussions. 
M.B.N.K. acknowledges support from the National Natural Science Foundation of China (grant 11573004). 
RS acknowledges support from the German Science Foundation (DFG) priority program SPP 1992 “Exploring the Diversity of Extrasolar Planets” under project Sp 345/22-1, and Yunnan Academician Workstation of Wang Jingxiu (No. 202005AF150025).
Xiaoying Pang acknowledges the financial support of the grant of National Natural Science Foundation of China, No: 12173029 and No. 12233013. 
This research was supported by the Postgraduate Research Scholarship (grant PGRS1906010) of Xi'an Jiaotong-Liverpool University (XJTLU). 
This paper utilizes data from the SVO Stars with debris disks and planets Data Access Service at CAB (CSIC-INTA) \footnote{\url{http://svocats.cab.inta-csic.es/debris2/index.php}}, catalog of resolved debris disks maintained by Nicole Pawellek and Alexander Krivov \footnote{\url{https://www.astro.uni-jena.de/index.php/theory/catalog-of-resolved-debris-disks.html}} and data from circumstellardisks.org \footnote{\url{https://www.circumstellardisks.org/}}.
This paper makes use of the \python{} packages NumPy \citep{numpy}, SciPy\footnote{\url{http://www.scipy.org}}, and Matplotlib \citep{mpl}.

\section*{Data Availability}

The data underlying this article will be shared on reasonable request to the corresponding author.

\bibliographystyle{mnras}
\bibliography{planet_on_comets}
\bsp	%
\label{lastpage}
\end{document}

%% file: sections/commands.tex
\newcommand{\msun}{$\mathrm{M}_\odot$}
\newcommand{\mdot}{$\mathrm{M}_\odot$}
\newcommand{\mj}{$\mathrm{M}_\mathrm{Jup}$}
\newcommand{\hmr}{$r_\mathrm{h}$}

\newcommand{\tms}{$t_\mathrm{ms}$}

\newcommand{\rtide}{$r_\mathrm{tide}$}
\newcommand{\rlagr}{$r_\mathrm{Lagr}$}
\newcommand{\rhill}{$r_\mathrm{Hill}$}

\newcommand{\mdotin}{\mathrm{M}_\odot}

\newcommand{\hmrin}{r_\mathrm{h}}

\newcommand{\tcrin}{t_\mathrm{cr}}
\newcommand{\trhin}{t_\mathrm{rh}}
\newcommand{\tmsin}{t_\mathrm{ms}}

\newcommand{\rtidein}{r_\mathrm{tide}}

\newcommand{\rhillin}{r_\mathrm{Hill}}

\newcommand{\mcl}{\texttt{MCLUSTER}}
\newcommand{\reb}{\texttt{REBOUND}}
\newcommand{\ias}{\texttt{IAS15}}

\newcommand{\nbo}{\texttt{NBODY6++GPU}}

\newcommand{\lps}{\texttt{LonelyPlanets}}
\newcommand{\amuse}{\texttt{AMUSE}}
\newcommand{\fortran}{\texttt{FORTRAN}}

\newcommand{\python}{\texttt{Python}}

\newcommand{\ptbs}{perturbers}
\newcommand{\ptb}{perturber}

\newcommand{\pdp}{test particle}
\newcommand{\pdps}{test~particles}

\newcommand{\Pdps}{Test particles}

\newcommand{\ensemble}{for the ensemble of all planetary systems in the star cluster}

\newcommand{\nbody}{{\it{N}}-body}
\newcommand{\fifju}{50-au-Jupiter}

%% file: sections/intro.tex
\section{Introduction}
\label{intro.sec}

Exoplanet surveys have provided a wealth of information on the formation and evolution of planetary systems in the Milky Way. To date, a total of 5365 exoplanets have been discovered\footnote{\url{http://exoplanet.eu/}; accessed on April 2023.}, of which around 30 have been found in star clusters \citep[see, e.g., table~1 of][for a recent review]{cai2019}. Despite the small fraction of known exoplanets in star clusters to date, it cannot be ruled out that exoplanets are common in star clusters. The origin of this dearth of known exoplanets in star clusters is still poorly understood, but it may be a combination of the following: (i) exoplanets may be less likely to form in star clusters; (ii) exoplanet systems may be disrupted following dynamical encounters with neighboring stars; or (iii) observational selection effects prevent the detection of exoplanets \citep[see, e.g.,][sec.~1 for a discussion]{cai2017, stock2020}.

Most protoplanetary disks evolve into planetary systems, containing a debris disk and/or planets. Debris disks are commonly found around main-sequence stars of all ages \citep{2018rev}. Observed outer disk structures can be explained by the presence of one or more planets in a system. The characteristics of systems in which debris disks and planets coexist are still under debate. For example, whether the presence of debris disks is correlated with planetary masses or planetary multiplicity, is still inconclusive (see, e.g., observations by \citealt{bryden2009, kospal2013, marshall2014, moro2015, maldonado2017}, and predictions by \citealt{raymond2012}). The outer regions of planetary systems are shaped by the surrounding stellar population, and are particularly vulnerable to external perturbations, such as flybys of neighboring stars and molecular clouds, and to some degree also by the Galactic tidal field.

Debris structures have been detected around a large number of stars\footnote{At the time of writing, 175 resolved debris disks have been reported; https://www.astro.uni-jena.de/index.php/theory/catalog-of-resolved-debris-disks.html ; accessed on April 2023.}, with typical distance to their host stars within hundreds of au, while a few extend to thousands of au (e.g., HD\,278932 and HD\,189002 \citealt{2021MNRAS.501.6168M}). The most notable example of a debris disk is that of our own Solar system, which hosts the main asteroid belt and the Kuiper belt. Some stars with debris disks in the solar neighborhood are known to host planets, such as $\epsilon$~Eri \citep{1998ApJ...506L.133G}, $\tau$~Ceti \citep{2004MNRAS.351L..54G}, 82~Eri \citep{2012A&A...548A..86L}, GJ~581 \citep{2015MNRAS.449.3121K}, Fomalhaut~A \citep{1998Natur.392..788H}. Although among all detected debris disks, however, there are only several dozens of systems in which planets and debris disks are known to coexist. Known debris structures among stars in star clusters are rare \citep{2011MNRAS.411.2186S, 2011MNRAS.413.1024P,2012ApJ...750...98U, 2012MNRAS.420.2884S, espaillat2017, 2020A&A...641A.156M}, but can provide valuable insights in the formation and dynamics of planetary systems in these environments. This can partially be attributed to observational biases, and however, may also be attributable to the role of environmental factors in the dynamical evolution of such debris structures.

Theories on the formation and evolution of planetary systems suggest that the properties of planets and debris disks are linked. Many observed structures of debris disks can indeed be explained by the presence of planets. It is likely that many of such debris disks are not remnants of protoplanetary disks or transition disks, but are instead products of collisions of larger objects (collisional cascade). The presence of a debris disk suggests the presence of objects sized 100~km or larger. On the other hand, kilometer-sized objects can under some circumstances contribute to the formation of planets \citep{2018rev}. It is therefore likely that when debris structures are observed, one or more planets may be present.

However, there are only a few known systems in which planets and debris disks are known to coexist, possibly because close stellar encounters have influenced the evolution of these systems at early times. Both stars, planets, and debris disks are thought to spend their infancy together in dense stellar environments, where close encounters with neighboring stars strongly affect the stability of planetary systems. Earlier studies on the dynamical evolution of planetary systems in star clusters suggest that planets may not be able to survive the impact of close encounters \citep[][and many others]{malmberg2007, spurzem2009, malmberg2011, liu2013configurations, hao2013, zheng2015, shara2016, cai2017, cai2019, ht2017, fran2019, li2019, van2019}. Numerical studies have suggested that close encounters can induce instabilities in disks and truncate disks, although part of such disks may still survive (see \citealt{2022review_single_encounter} and references therein for studies on single encounters, and see, e.g., \citealt{hands2019, veras2020} for studies on disks within star cluster simulations).

Apart from the stellar environment, planets may also contribute to shaping debris disks \citep{2018rev}. When this interaction is sufficiently strong, the distribution of the debris particles may be used to constrain the properties of an unseen planet \citep{tabeshian2016,tabeshian2017}. A similar approach was taken in attempts to locate the hypothetical "Planet 9" in the outer regions of the Solar system \citep[see, e.g.,][and references therein]{2014MNRAS.444L..78I,2014Natur.507..471T,2015MNRAS.453.3157J,2016MNRAS.457L..89M}.

In this study, we carry out \nbody{} simulations to explore the evolution of debris disks in star clusters, after gas has been removed from the star cluster and the circumstellar gas has dissipated. The study focuses on the influence of planets on debris disks, in an attempt to further study why debris disks, and debris disks with planets, are rarely observed. By evolving such systems in dense stellar environments, the birthplaces of stars, planets, and debris disks, we help deepen our understanding of the rareness of debris disks observed in star clusters and the scarcity of known planetary systems where planets and debris disks co-exist. This paper is organized as follows. In Section~\ref{section:method} we describe the initial conditions and numerical approach. We present the results in Section~\ref{section:results}. Finally, we summarize and discuss our conclusions in Section~\ref{section:conclusions}.

%% file: sections/method.tex
\section{Numerical Methods and initial conditions}
\label{section:method}

\subsection{Computational challenges}
\label{section:comp_challenge}

\begin{figure}
    \centering
    \includegraphics[width=\columnwidth]{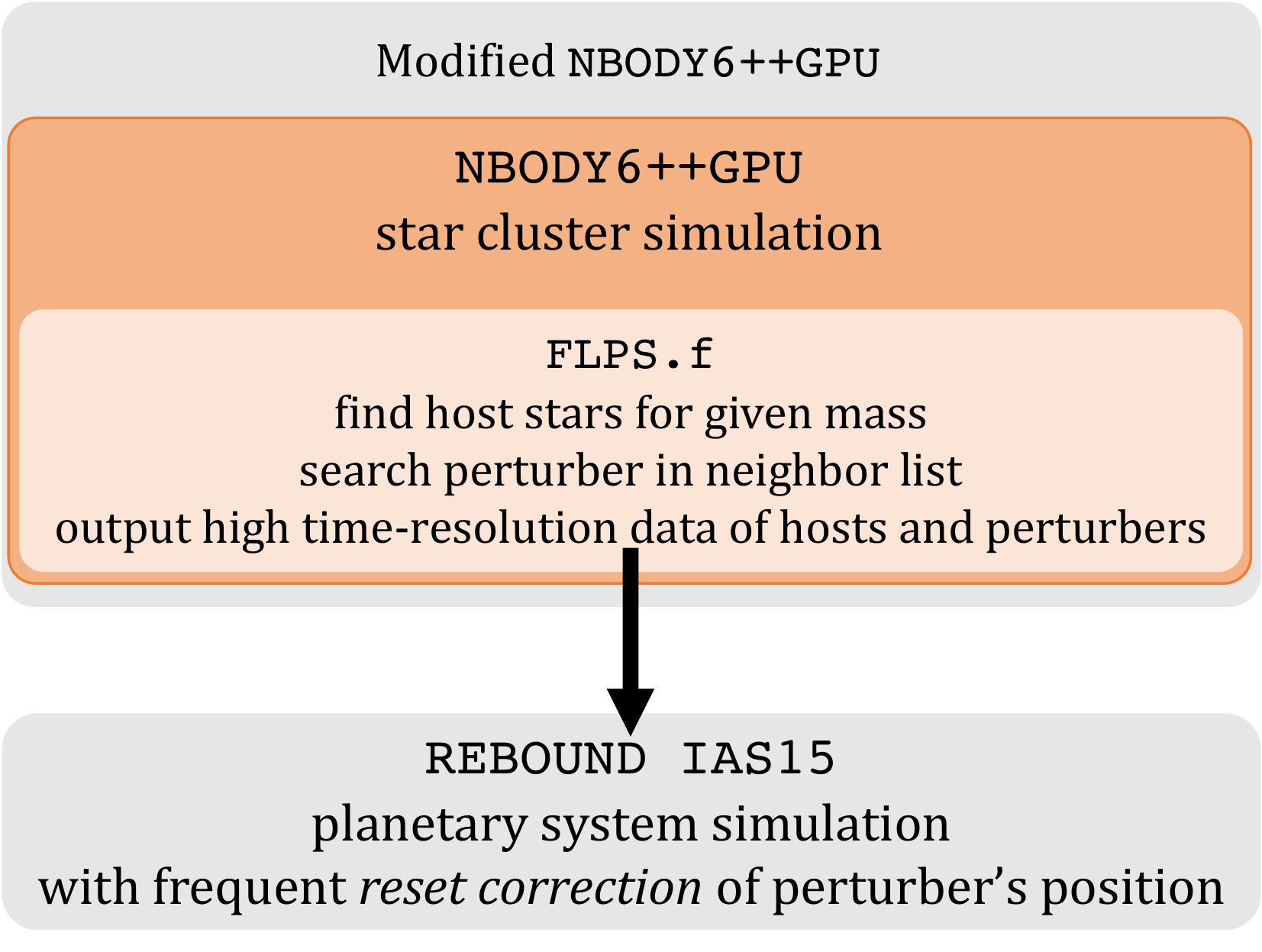}
    \caption{Schematic diagram of the computational approach adopted in this study.}
    \label{fig.schem}
\end{figure}

A major challenge in simulating the dynamical evolution of planetary systems in star clusters originates from the vast range in spatial and temporal scales of these systems. The integration time step required for planetary systems is many orders of magnitude smaller than that of the stellar population. Another challenge arises from the number of particles in a planetary system. For realistic systems, the total number of planetary debris particles greatly outnumbers the stars in the simulation. Given that the time complexity of \nbody{} simulators typically ranges between $\mathcal{O}(n\log{}n)$ and $\mathcal{O}(n^2)$, this can pose a significant challenge when attempting to integrate such systems.

Since planets and debris particles have a negligible gravitational influence on the stellar population, it is possible to bridge star cluster simulations with simulations of planetary systems. The stellar population can be integrated first, and the dynamics of the planetary systems can be computed afterward, using the kinematic data obtained for the stellar population. This approach was taken in \cite{cai2017}, who developed the code \lps{} in the \amuse{} framework \citep{amuse}. This code first uses \nbo{} \citep{NBODY6++GPU} to integrate the stellar population, and subsequently \reb{} \citep{reb} to integrate the planetary systems.

To complete the simulations within reasonable wall-clock time, choices of the following technical details may affect the accuracy when using the \lps{} scheme or similar schemes:

\begin{enumerate}[leftmargin=0.6cm, labelsep=3pt, itemindent=-5pt]

    \item The number of \ptbs{}. In \lps{}, the influence of the star cluster on the planetary systems is modeled by integrating the orbits of the nearest stellar neighbor(s) of the planetary system. \citet{cai2017}'s method uses one neighbor star to represent the perturbations of the star cluster. This is in many cases reasonable, as the tidal force experienced by the planetary system is dominated by that induced by the nearest neighbor star. \citet{cai2018} and \citet{fran2019} follow this method and also adopt one stellar neighbor. \citet{cai2019} performed a convergence test on the number of perturbers, and their result generally converges when including five or more perturbers. The choice of five \ptbs{} is also adopted in \citet{stock2020}. %
    \item The output frequency. Accurate integration of the planetary systems in \reb{} requires \nbo{} to store the kinematic data much more frequently than normally required for star cluster simulations. Storing data at a high time resolution slows down \nbo{} significantly, and requires much storage space. \citet{stock2020} chose 10000~years as the output interval of star cluster simulation to balance accuracy and speed. 
    \item The \ptbs{}' orbits. Another often overlooked issue is that the \ptbs{} are also subject to the gravitational interaction with their own neighbors (stars that affect the orbits of \ptbs{}), and these interactions are sometimes neglected in simulations of planetary systems, where, as a consequence, the trajectories of \ptbs{} may deviate. Since dynamical motions of \ptbs{} can impact the fate of the planetary systems, it is possible that a small deviation in trajectories without correction may greatly change the results of planetary system simulations.  
        
\end{enumerate}
Our scheme (Section~\ref{section:computational}) is similar to that of \lps{} in modelling the perturbation of star cluster members on planetary systems, but we optimize the code for accuracy and speed.

\subsection{Computational approach} 
\label{section:computational}

We select the ten nearest neighbor stars as perturbers, to obtain precise modeling of the star cluster environment. This choice does not significantly increase the simulation time, because (i) the simulation time is dominated by a large number of particles in planetary systems (Section~\ref{sec.plsysini}), and (ii) the integration time step is dominated by the innermost orbits, which is not influenced by distant \ptbs{}.

To overcome the time-resolution difficulty, we have developed a \fortran{} routine \texttt{FLPS.f} for the latest version of \nbo{}\footnote{The latest version of \nbo{} with long-term support is at \url{https://github.com/nbody6ppgpu/Nbody6PPGPU-beijing}.} \citep{NBODY6++GPUNewSSE}, that allows storing kinematic data at a higher frequency without a significant sacrifice of simulation speed (see Figure~\ref{fig.schem}). This routine identifies perturbers in the neighbor list of each star in \nbo{}, rather than searching the entire cluster. It stores only information required for the planetary system simulation: masses, positions, and velocities of host stars and their perturbers. This greatly reduces storage requirements and speeds up the simulation. Besides the regular output at fixed intervals, we also store the data when the host star's neighbor list changes. The median output interval of the host and \ptb{} data is approximately 50~years.

In order to model accurate trajectories of \ptbs{}, we perform the planetary system simulations including \ptbs{} using {\em reset-correction} approach. First, we simulate the system for short time intervals (the {\em reset-time}), so that \ptbs{} will not deviate too far from their trajectories in the star cluster. Second, we start a new \reb{} simulation instance using the kinematic data of the host star, planets, and \pdps{} from the previous simulation, and the \ptbs{} with their correct kinematic data from the star cluster simulation. We repeat these two steps until the desired simulation time is reached. If we choose the {\em reset-time} small enough, the motion of \ptbs{} is modeled correctly when integrating the planetary systems. In this work we adopt a {\em reset-time} of 100~years, to balance the costs of accuracy and wall-clock time. We implement the {\em reset-correction} method with \reb{} in \texttt{C} code, using the \ias{} integrator \citep{ias15}.

\subsection{Initial conditions: star cluster}

\begin{table}
    \caption{Initial conditions of the star cluster model.}
    \label{tab.scini}
    \centering
    \begin{tabular}{ll}
        \hline \hline
        Quantity & Value \\
        \hline
        Number of stars           & $N=8000$        \\
        Total cluster mass        & $M=4662$~\mdot{}    \\
        Half-mass radius          & $\hmrin{}=0.78$~pc       \\
        Density profile           & \citet{plum1911}   \\
        Initial mass function     & \citet{kroupa2002}; $0.08 - 150$~\mdot{}  \\
        Virial ratio              & $Q=0.5$     \\
        Tidal field               & Solar orbit in the Milky Way \\
        Tidal radius              & $\rtidein{}=23.62$~pc       \\
        Primordial binaries       & none        \\
        Metallicity               & $Z=0.001$  \\
        Stellar evolution         & enabled      \\
        Integration time          & 100~Myr \\
        \hline \hline
    \end{tabular}
\end{table}

We model the evolution of planetary systems in an intermediate-mass star cluster. To better compare our results with previous studies, we adopt initial conditions similar to the 8k star cluster model in \citet{stock2020}, generated by \mcl{} \citep{mcluster}. The initial conditions are summarized in Table~\ref{tab.scini}. Initial stellar positions and velocities are generated from the \citet{plum1911} model, with initial stellar masses drawn from the \citet{kroupa2001a} IMF, in the mass range $0.08-150$~\mdot{}. The star cluster is initially in virial equilibrium, and evolves in solar orbit in a Milky Way tidal field. We do not include primordial binary stars. All stars are initialized with a metallicity of $Z=0.001$. Stellar evolution in \nbo{} including stellar winds is enabled with the implementation of \citet{NBODY6++GPUNewSSE}, which is an improvement of the original algorithms of \citet{hurley2000-sse}. Following the analysis of \citet{cai2019}, \citet{veras2020}, and \citet{stock2020}, we integrate the planetary systems for 100~Myr.

\subsection{Initial conditions: planetary systems}
\label{sec.plsysini}

\begin{figure*}
    \centering
    \includegraphics[width=2\columnwidth]{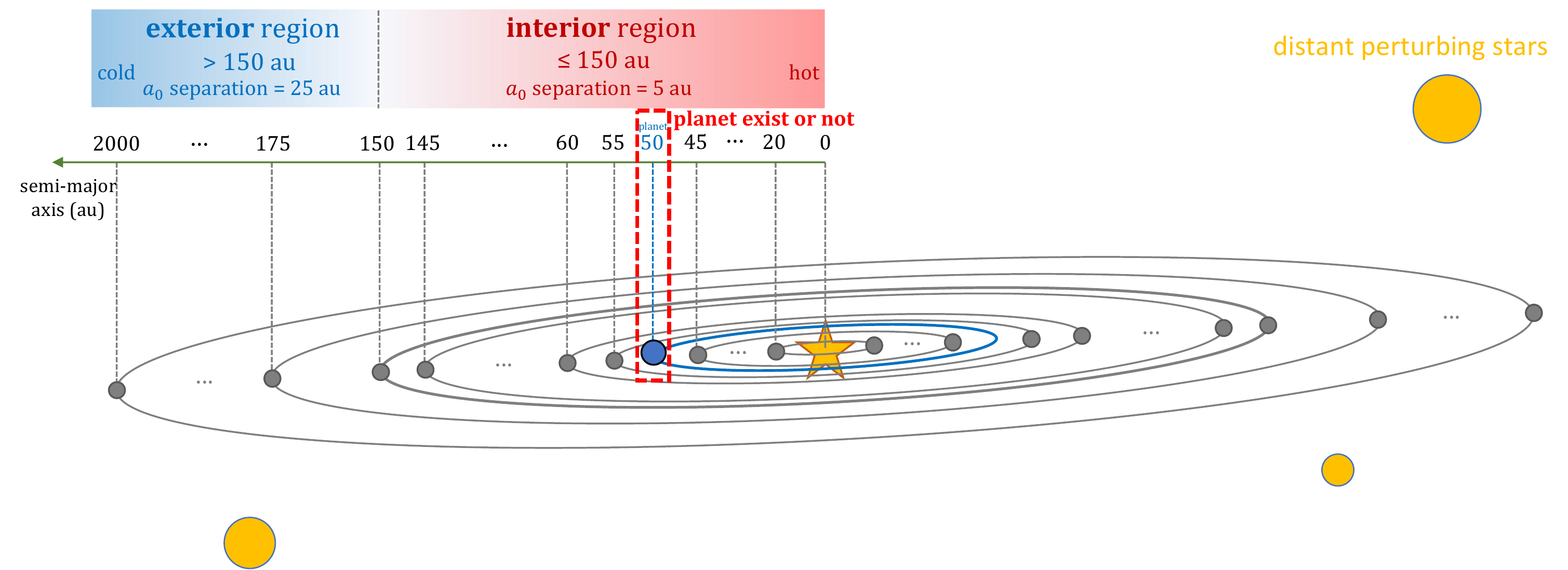}
    \caption{Schematic diagram of initial conditions of the planetary systems. Each planetary system contains a "\fifju{}" and 200 \pdps{}. We also include ten nearest neighbor stars (not fully shown here) in the simulation.}
    \label{fig.plsysini}
\end{figure*}

\begin{table}
    \caption{Initial conditions of the planetary systems. Each simulation contains a hundred planetary systems in which the host stars have masses of $M\approx 1$\,\mdot. We include ten perturbers when modeling the planetary systems. Planets and \pdps{} are removed when the distance to their host star exceeds 5000~au.}
    \label{tab.plini}
    \begin{tabular}{lll}
        \hline \hline
        Property               & Planet      & \pdps{}     \\
        \hline                                                     
        Number                 & 0 or 1       & 200        \\
        Mass                   & $m_p=317.83 M_\oplus$ ($M_\mathrm{J}$)  & $m_c=0$       \\
        Semi-major axis        & $a_p=50$~au     & $a_c=20-2000$~au    \\
        Eccentricity           & $e_p= 0.0484$ ($e_\mathrm{J}$)   & $e_c=0.01$             \\
        Inclination            & $i_p= 0^\circ$ (reference plane)  & $i_c = 0.01~\mathrm{rad}=0.573^\circ$ \\
        \hline \hline
    \end{tabular}
\end{table}

In our study, we focus on planetary systems hosted by solar-type stars. We select the 100 stars with masses closest to 1~\msun{} for further analysis. The resulting mass range of these host stars is $0.981-1.015$~\mdot{}. Each of the 100 host stars is assigned 200 \pdps{}. Figure~\ref{fig.plsysini} provides a schematic view of the structure of the planetary systems.

To study the effect of the planet on \pdps{}, we carry out two sets of simulations: one set in which a planet (a "\fifju{}") is present, and one set in which the planet is absent. The other initial conditions of the two sets are identical. The planet has a mass and an eccentricity equal to that of Jupiter, because we expect the most massive planet among the ones that we are familiar with can bring us statistical significance on its impact on \pdps{}. Inspired by \citet{nesvold2015}, we initialize its initial semi-major axis to 50~au. The initial orbit of the planet defines the reference plane for each planetary system, to which the orbital configuration at later stages will be compared.

We initialize the \pdps{} over a wide range of semi-major axes. Inspired by \citet{veras2020}, we classify \pdps{} according to their semi-major axis: \textit{interior} region means $a \le 150~\mathrm{au}$, and \textit{exterior} region means $a > 150~\mathrm{au}$. Interior \pdps{} are placed at intervals of 5~au (except 50~au, the location of the planet) from 20~au to 150~au. Exterior \pdps{} are placed at intervals of 25~au from 175~au to 2000~au. At each of the 100 semi-major axes, two \pdps{} are initialized in symmetrical positions about the star. All \pdps{} initially have co-planar orbits with eccentricity $e_c=0.01$ and inclination of $i_c = 0.01~\mathrm{rad} = 0.573^\circ$. They are set as type 0 test particles (massless) in \reb{}, so that they do not exert force on other bodies, but they experience the gravitational force from the host star, the planet, and \ptbs{}. The initial argument of periastron and longitude of ascending node are randomized.

The masses of the host stars ($\sim 1\,M_\odot$) are not affected by stellar evolution on the timescale of the simulations, while the masses of more massive cluster members evolve significantly over time. In the planetary system simulations (\reb) we use the updated masses from \nbo{} for the stellar neighbors, and as such we include the dynamical effects resulting from stellar evolution.

The small size of \pdps{} and sparse distribution rarely result in physical collisions. \citet{nesvold2017} included 10,000 particles in the range 65-85~au from a 2.5~\msun{} star with an 11~\mj{} perturber at 650~au, and they show that the effect of collisions on eccentricity is less than 10\%, and even less for inclination, longitude of nodes and argument of pericenter. The typical distance between the particles in our simulations is substantially larger than in those of \citet{nesvold2017}, so collisions are rare and are thus not implemented in this study.

Under the influence of internal scattering events and external perturbation, a particle (planet or \pdp{}) may escape from its planetary system. A particle is labeled as having escaped from its host star when its distance to the host star is larger than $5,000$~au, which is typically half the initial distance between the host star and the nearest neighbor star. Upon escape, the particle is removed from the \reb{} simulation.

We include a Galactic tidal field when simulating the star clusters, but we exclude it from the planetary system simulations. Outer regions of planetary systems, including the solar system's Oort cloud, are affected by the Galactic tidal field \citep{1993rev}, which can alter orbits and strip comets from the outer regions. The tidal radius around a solar-mass star in the simulated environment (i.e., the solar neighborhood) is 
\begin{equation}
    R_{\mathrm{tide}} \approx a_\odot 
    \left( {\frac{m_\odot}{3 M_\mathrm{enclosed}}} \right)^{1/3}
    \approx 1.7~\mathrm{pc}~
    ~\quad ,
\end{equation}
which is substantially larger than our criterion used for removing escaping particles (5000~au). The typical distance between the nearest neighbor and the host star is of order $10^5$~au. (see Section~\ref{section:cluster_result}). The effect of the Galactic tidal field can thus be ignored when modeling the planetary systems.

Table~\ref{tab.plini} lists the initial conditions of the planetary systems. By comparing systems with and without a planet, we are able to constrain the influence of the planet. This is nearly true, although minor differences occur due to (i) round-off errors, and (ii) the enclosed mass within a \pdp{}'s orbit increases by $\sim 0.1\%$ due to the planet. These differences can lead to deviations on a long timescale; we will discuss this in Section~\ref{theory.sec}.

%% file: sections/result.tex
\section{Results}
\label{section:results}

\subsection{Isolated planetary systems}
\label{theory.sec}

\begin{figure}
    \includegraphics[width=\columnwidth]{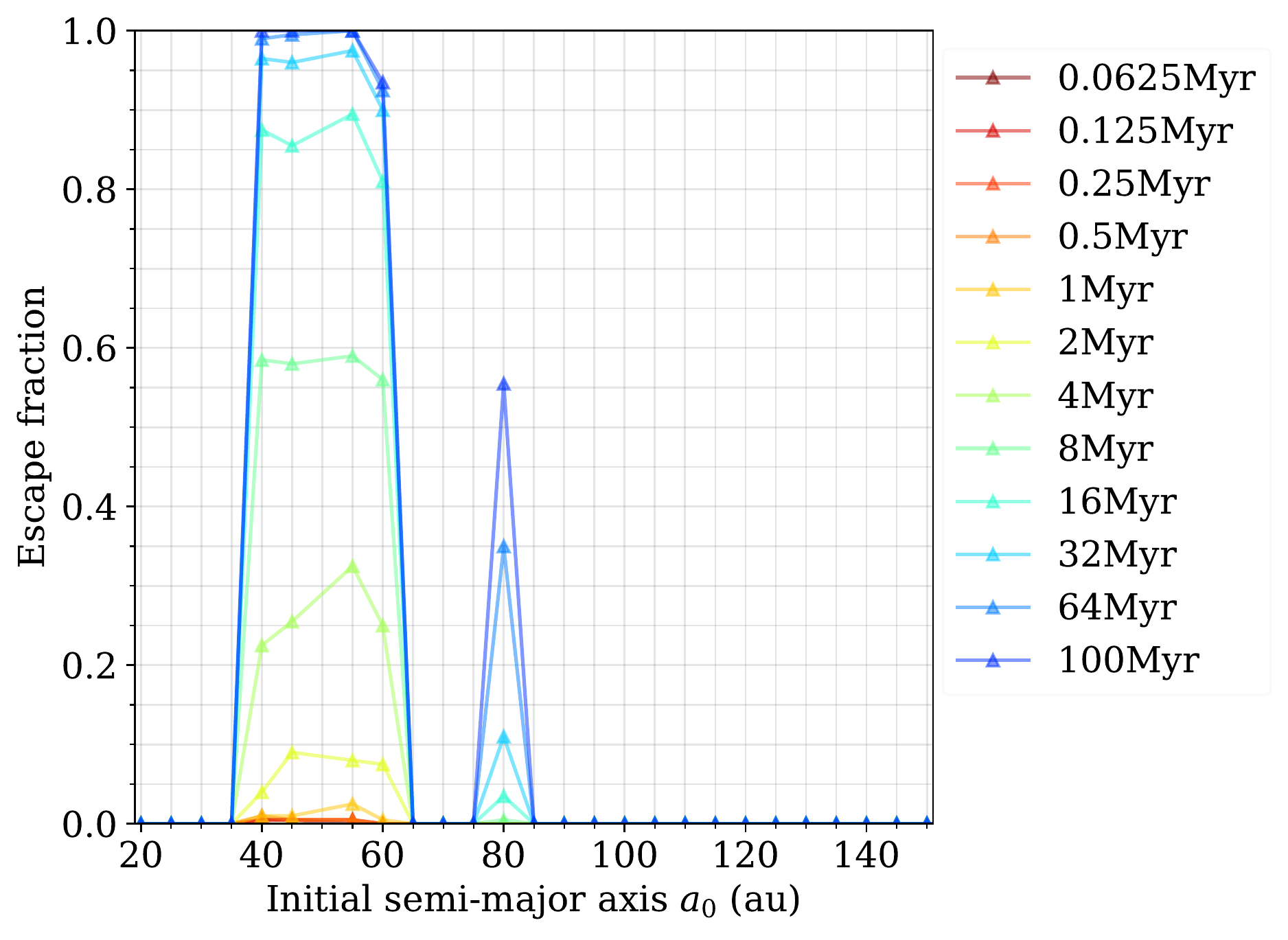} 
    \caption{Escape fraction of \pdps{} as a function of the initial semi-major axis at different simulation times, for the isolated planetary system.} 
    \label{fig.iso_esc_fraction_over_a0_by_t_x200} 
\end{figure}

\begin{figure}
    \includegraphics[width=\columnwidth]{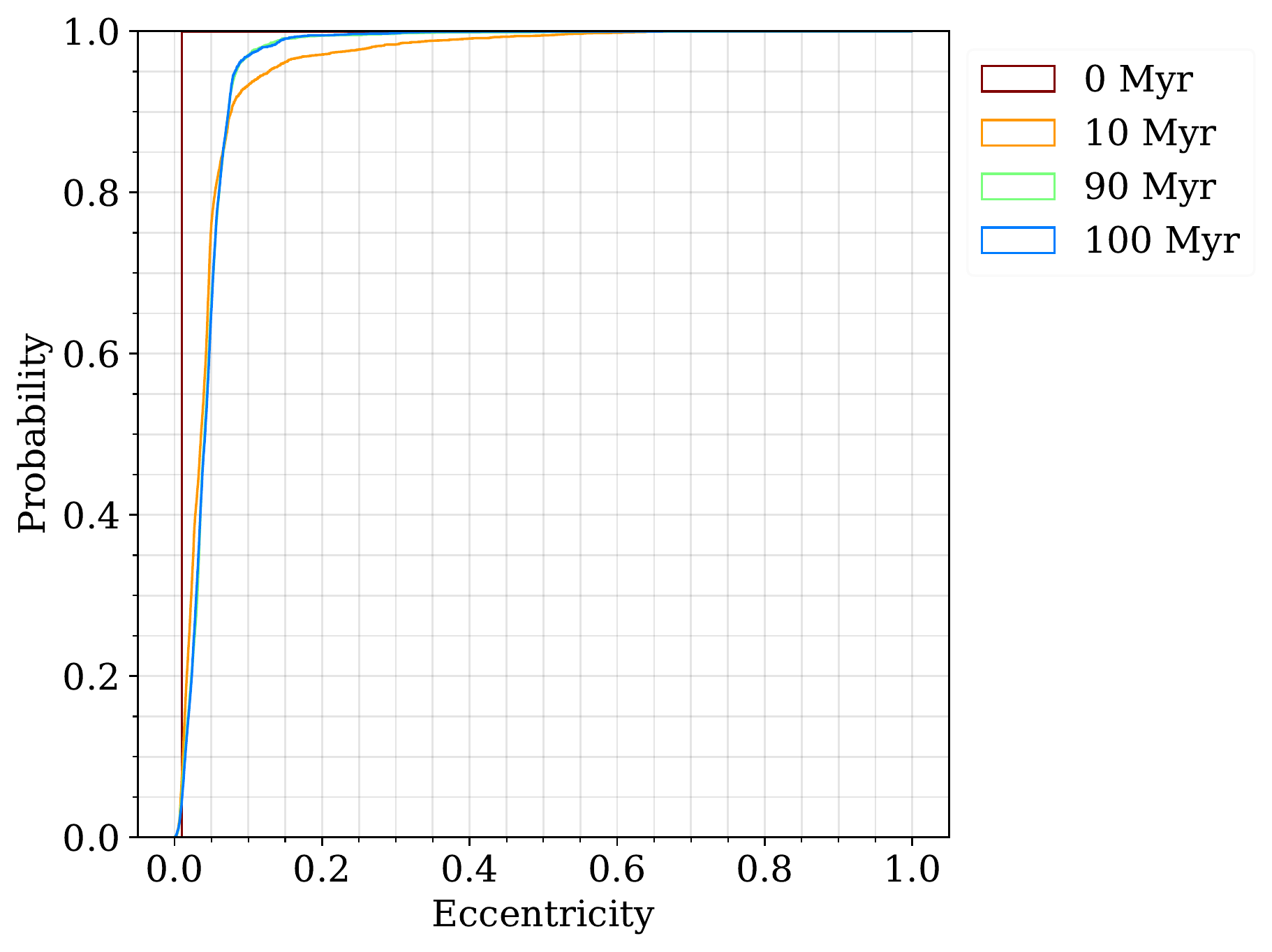} 
    \caption{Cumulative eccentricity distribution of surviving \pdps{} in the interior region ($a \le 150$~au).} 
    \label{fig.iso_CDF_e_in} 
\end{figure}

To identify the influence of the neighboring stellar population on planetary systems, we first briefly review the evolution of isolated planetary systems (a star, a planet, and \pdps{}). Detailed studies about isolated planet-hosting systems with debris disks can be found in \citet{nesvold2015, Nesvold2016, tabeshian2016, tabeshian2017}. 

Since \pdps{} are massless and do not interact with each other, the planetary system behaves as a collection of separate three-body systems consisting of a star of mass $M$, a planet of mass $m_p$, and a massless particle. Such systems have been studied extensively \citep[see, e.g.,][for an extensive discussion on this topic]{threebodyproblembook}. The degree of stability of such three-body systems can be, to first order, quantified using the Hill radius, \rhill{} of the planet, 
\begin{equation}
    \rhillin \approx a_p (1-e_p) (\frac{m_p}{3 M})^{\frac{1}{3}}
    ~\quad,
\end{equation}
where $a_p$ and $e_p$ are the semi-major axis and eccentricity of the planet respectively. The system has a high degree of stability when \pdps{} remain at distances much larger than \rhill{} from the planet. In our models, $M \approx 1$~\msun{}, $m_p = 1$~M$_{J}$, $a_p=50$~au, and the planet's orbit is nearly circular, so the Hill radius is $\rhillin{} \approx 3.25 $~au. \Pdps{} on circular orbits with semi-major axes in the range $a_p-\rhillin{} \lesssim a_c \lesssim a_p+\rhillin{}$ are almost instantly scattered away from their orbits. As time passes, the affected region broadens \citep[see, e.g.,][for a discussion on this topic]{nesvold2015}.

To further analyze the evolution of isolated planetary systems, we carry out additional simulations. Each system consists of a solar-mass host star, 200 \pdps{}, zero or one \fifju{}. Stellar neighbors are not included at this stage. Figure~\ref{fig.iso_esc_fraction_over_a0_by_t_x200} shows the escape fraction of \pdps{} as a function of the initial semi-major axis at different times. The planet at 50~au ejects nearly all \pdps{} from 40~au to 60~au, and also removes more than half of the \pdps{} that are initially located at 80~au. The other particles do not escape, while the \pdps{} in the interior region ($a \le 150$~au) gain higher eccentricities (Figure~\ref{fig.iso_CDF_e_in}). The inclination for interior particles, and semi-major axes, eccentricities, and inclinations for exterior particles show nearly no change.

\subsection{Dynamical evolution of the star cluster}
\label{section:cluster_result}

\begin{figure}
    \includegraphics[width=\columnwidth]{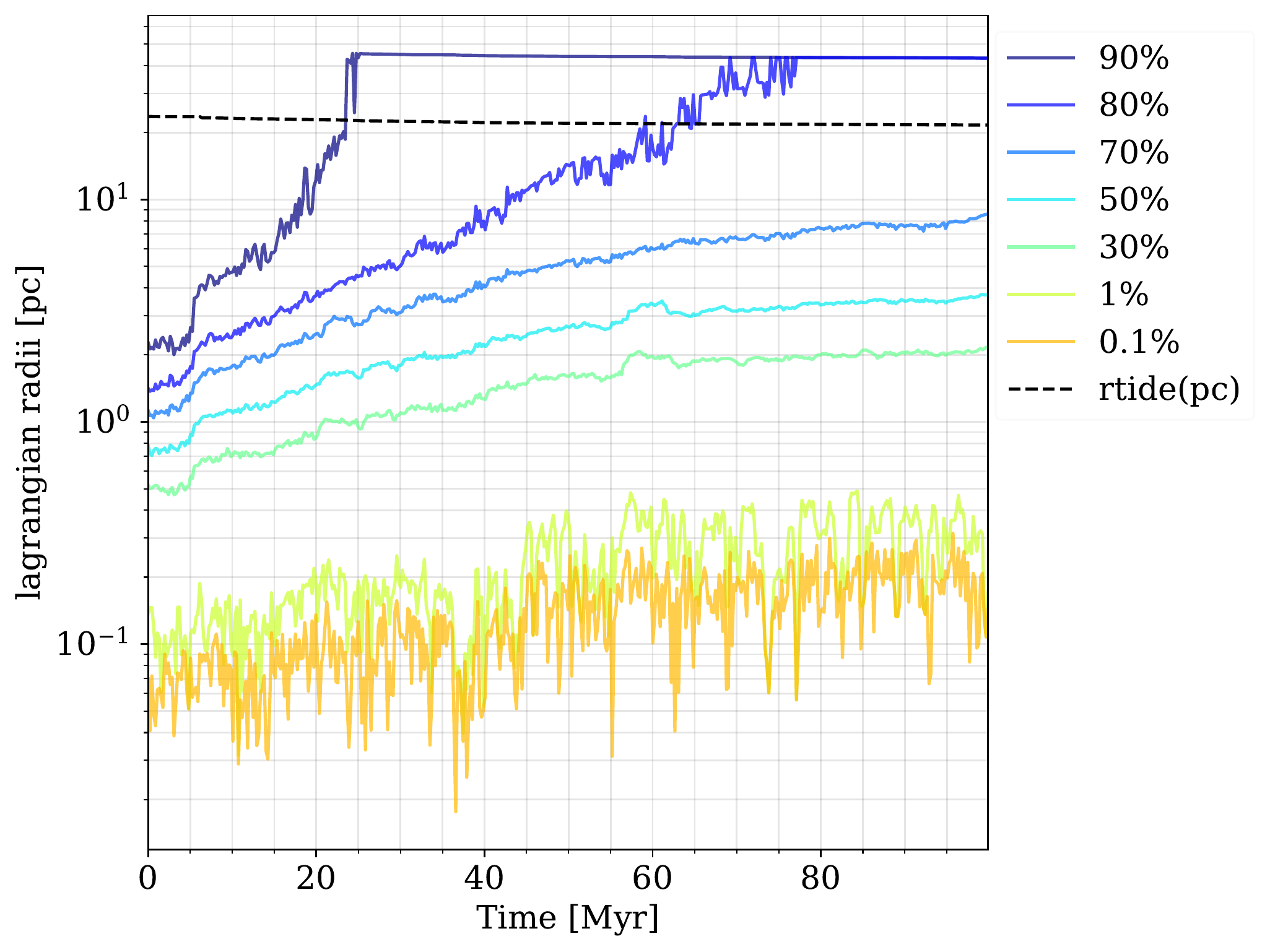}
    \caption{Evolution of the Lagrangian radii \rlagr{} of the star cluster, containing the fraction of initial total mass. The dashed black curve indicates the tidal radius of the cluster, which gradually decreases from 23.62~pc to 21.62~pc (a reduction of 8.4\%) as the star cluster loses mass. }
    \label{fig.8k_rlagr} 
\end{figure}

\begin{figure}
    \includegraphics[width=\columnwidth]{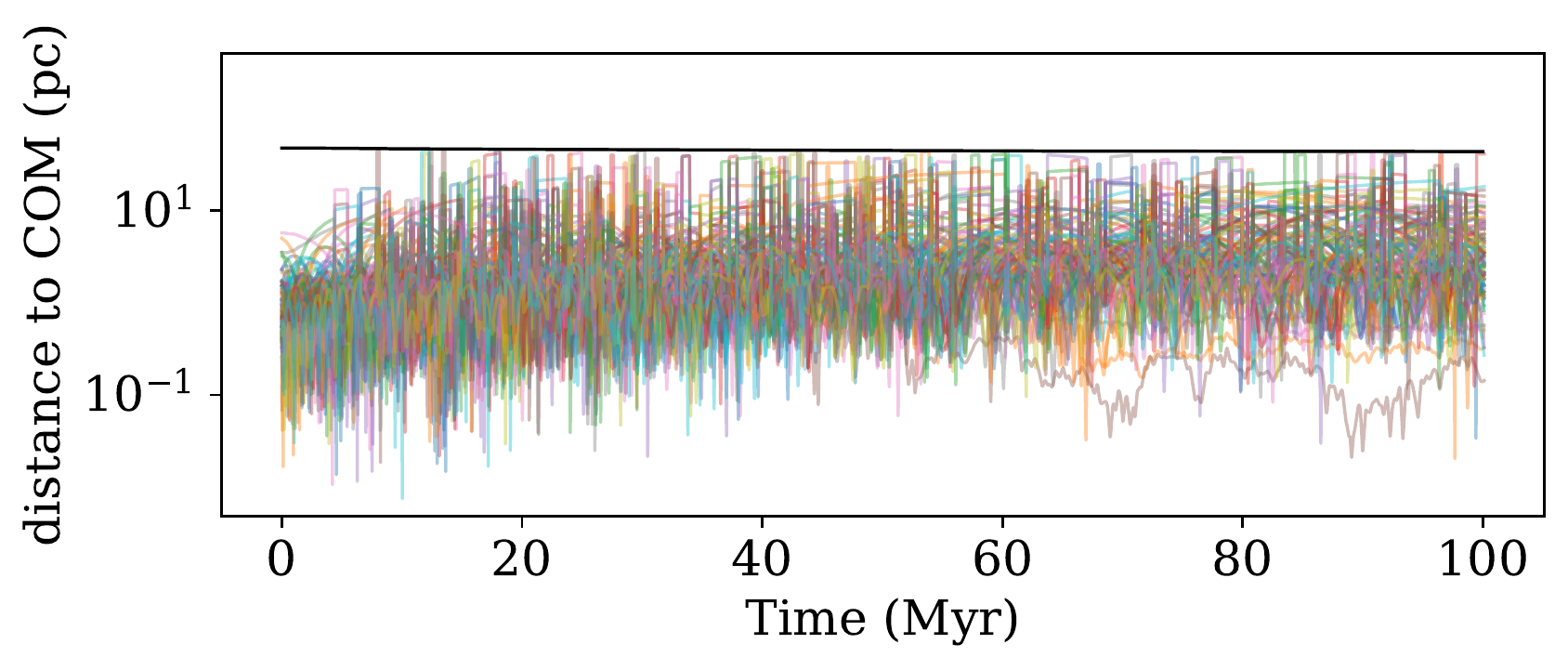}
    \caption{Distance of host stars to the center-of-mass of the star cluster, as a function of time. The black curve is two times of the tidal radius (2\rtide{}), which shrinks very slowly with time.}
    \label{fig.host_distance} 
\end{figure}

\begin{figure}
    \includegraphics[width=\columnwidth]{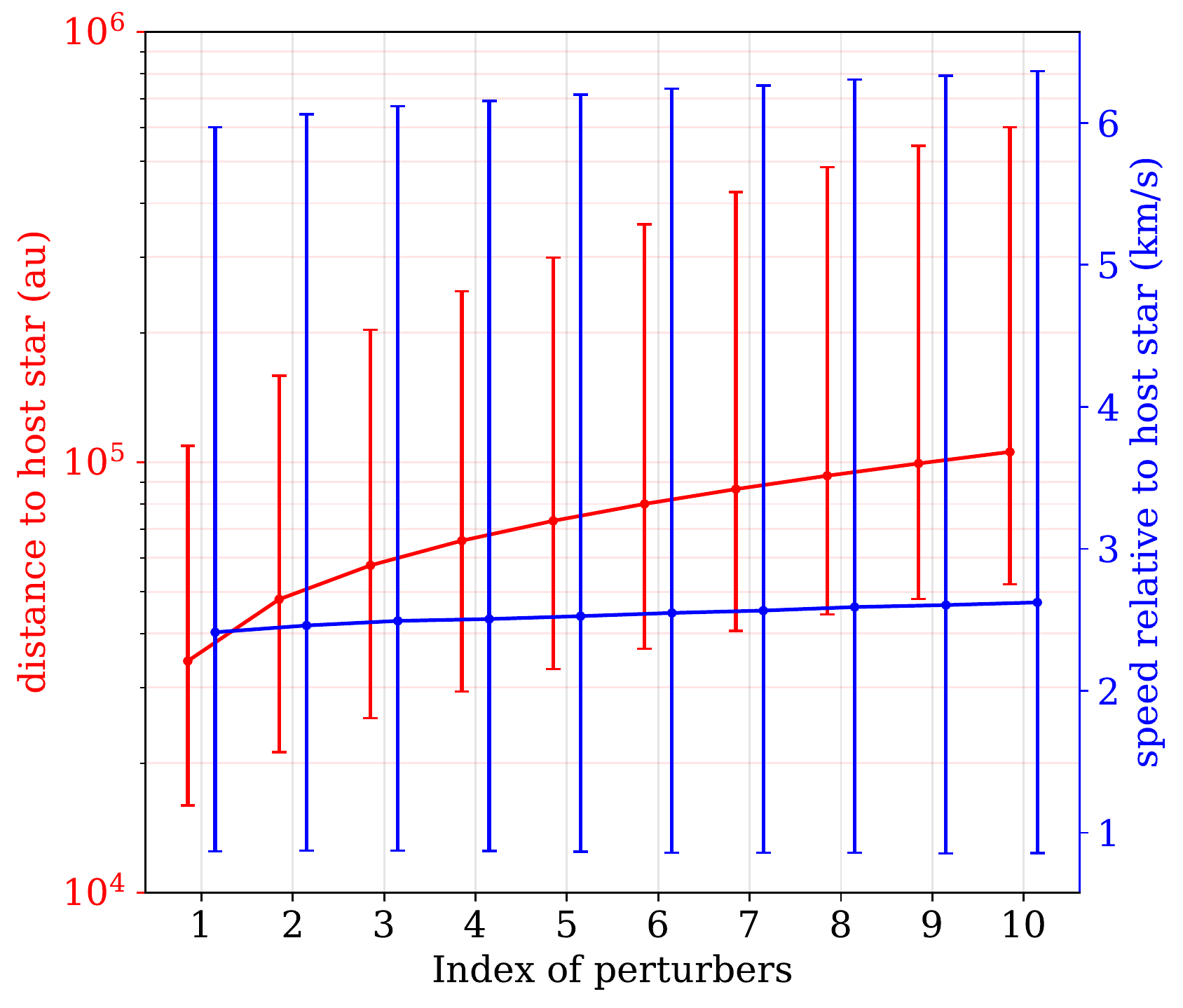}
    \caption{Distance and speed of \ptbs{} relative to the host star, \ensemble{}, for all simulation times combined. The solid curves connect the median values, while the error bars represent the first quartile (25\%) and the third quartile (75\%). }
    \label{fig.ptb_distance_speed} 
\end{figure}

\begin{table}
    \caption{Evolution of star cluster parameters that impact the constituent planetary systems. \hmr{}: half-mass radius. $\tau_{\mathrm{enc}}$: encounter timescale. $\rho_*$: number density. $\rho$: mass density. $D$: typical distance between two stars.}
    \label{tab.sc_evo}
    \centering
    \begin{tabular}{cccccc}
        \hline
        Time  & \hmr{} & $\tau_{\mathrm{enc}}$ & $\rho_*$             & $\rho$   & $D$    \\ 
        (Myr) & (pc)   & (Myr)                 & ($\mathrm{pc}^{-3}$) & ($\mathrm{M}_\odot~\mathrm{pc}^{-3}$)  & (pc)  \\
        \hline
        0    & 0.78     &  1.11   & 2186     & 1219     & 0.1    \\
        100  & 2.83     &  26.5   & 31.55    & 19.55    & 0.4    \\
        \hline        
    \end{tabular}
\end{table}

The properties of the surrounding stellar environment evolve over time, and therefore also the frequency and impact of close encounters with the planetary systems. Figure~\ref{fig.8k_rlagr} shows the evolution of the Lagrangian radii of the star cluster. By $t=100$~Myr, 21.2\% (988.32\msun{}) of the initial mass (4661.9\msun{}) of the star cluster is lost. Among this, 5.24\% (244.18\msun{}, 387~stars) of initial mass was lost due to stars escaping from the star cluster. The remainder of the mass is lost due to stellar evolution. The half-mass radius of the cluster evolves from 0.78~pc at $t=0$~Myr to 2.83~pc at $t=100$~Myr. Some stars escape, but all planet-hosting stars remain part of the star cluster throughout the simulation. As shown in Figure~\ref{fig.host_distance}, none of the host stars reaches a distance beyond two tidal radii (the adopted escape criterion for our star cluster model).

The characteristic timescale of the stellar population helps us understand how star clusters evolve. We adopt the definitions of the crossing time ($\tcrin{}$), the half-mass relaxation time ($\trhin{}$) and the mass segregation timescale ($\tmsin{}$) used in previous studies \citep[e.g.][]{spitzer1987, lamers2005, binney2008, Khalisi2007}. For our the star cluster model the initial values are $\tcrin{} = 0.209$~Myr, $\trhin{} = 20.2$~Myr, and $\tmsin{} = 11.8$~Myr, where \tms{} is for calculated for solar-mass stars. Many other quantities can also be used to characterize how the star cluster properties impact planetary systems. The first is the close encounter timescale, as estimated in equation~(3) of \citet{malmberg2007}:
\begin{equation}
    \begin{aligned}
        \tau_{\mathrm{enc}} \simeq &~ 5 \times 10^{7}~\mathrm{yr}~\left(\frac{\bar{m}_{*}}{1 ~\mathrm{M}_{\odot}}\right)~\left(\frac{r_{\mathrm{h}}}{1~\mathrm{pc}}\right)^{5 / 2}~\left(\frac{100~\mathrm{M}_{\odot}}{m_{\mathrm{cl}}}\right)^{1 / 2} \\
        & \times\left(\frac{10^{3}~\mathrm{au}}{r_{\min }}\right)~\left(\frac{\mathrm{M}_{\odot}}{m_{\mathrm{t}}}\right)
        \quad.
    \end{aligned}
    \label{eqn.encounter_timescale}
\end{equation}
Here, the average stellar mass is $\bar{m}_{*}=0.583~\mdotin{}$, the half-mass radius is $\hmrin{}=0.78$~pc, and the total cluster mass is $m_\mathrm{cl}=4662~\mdotin{}$ at $t=0$~Myr. We find that $\tau_{\mathrm{enc}}\approx1.11$~Myr for $r_{\min}=2000$~au and $m_{\mathrm{t}}=1~\mdotin{}$ at $t=0$~Myr. Note that the star cluster becomes sparser over time, and that $\tau_{\mathrm{enc}}$ evolves to ${\sim}26.5$~Myr at $t=100$~Myr. We therefore expect a typical planetary system to experience ${\sim}10$ encounters within 2000~au during the 100~Myr simulation. Such encounters can strongly disturb the outer \pdps{} and expel most of these (see Section~\ref{section:esc_fraction}). The number of encounters that a planetary system experiences within a certain distance are roughly proportional to that distance. The average number of encounters within 50~au within a timespan of 100~Myr is 0.31. Such encounters may result in the excitation or ejection of \pdps{}. We will return to this issue in Section~\ref{section:planet}.

Table~\ref{tab.sc_evo} lists several properties of the stellar population surrounding the planetary systems, for $t=0$ and $t=100$~Myr. The average stellar number density within the half-mass radius is $\rho_* = N_\mathrm{h}/(\frac{4}{3}\pi R_\mathrm{h}^3)$, where $N_\mathrm{h}$ is the number of stars within the half-mass radius. The stellar mass density within the half-mass radius is $\rho = M_\mathrm{h}/(\frac{4}{3}\pi R_\mathrm{h}^3)$, where $M_\mathrm{h}$ is the mass within the half-mass radius. The typical distance between the stars within the half-mass radius is $D = 2 R_\mathrm{h} N_\mathrm{h}^{-1/3}$. The average stellar number density and mass density decrease by a factor of $60-70$ during the simulation. The influence of the stellar neighbors on the planetary system is thus expected to reduce significantly over time. This is consistent with the findings presented in Section~\ref{section:results}.

In Section~\ref{section:comp_challenge} we motivated our choice for including 10 \ptbs{} when simulating the planetary systems. The range in distances and speeds of these \ptbs{} are shown in Figure~\ref{fig.ptb_distance_speed}, and confirm that the choice of including 10 \ptbs{} is adequate. The 1st to 5th nearest \ptbs{} have a median distance between $3.5\times10^4 - 7.3\times10^4$~au, while the corresponding values for the 6th to 10th \ptbs{} are $8.0\times10^4 - 10.5\times10^4$~au. If we adopt $F\propto r^{-2}$ to represent the force of the \ptbs{} on planetary systems, the force of the latter 5 \ptbs{} is equivalent to $\sim$30\% of the former. The influence of distant \ptbs{} on planetary systems is tidal in nature, which has the proportionality $\Delta F \propto r^{-3}$. The tidal contribution of the 5 distant \ptbs{} is equivalent to 15\% of that of the nearest 5. Although the choice of 5 \ptbs{} adopted in previous studies thus also serves as a reasonable approximation, including 10 \ptbs{} gives more accurate results.

Our star cluster model is similar to the 8k model in \citet{stock2020}, but is slightly different in terms of the adopted IMF, and consequently the initial total mass. \citet{stock2020} present the evolution of the tidal radius in their figure~2, but do not remove escapers. In our simulation, on the other hand, escaping stars are removed from the star cluster when their distance to the cluster center exceeds 2~\rtide{}. For example, their 90\% Lagrangian radius evolves to $\sim$2000~pc as stars escape the cluster, while in our simulation it remains within 10~pc.

\subsection{The planet} \label{section:planet}

\begin{figure}
    \includegraphics[width=\columnwidth]{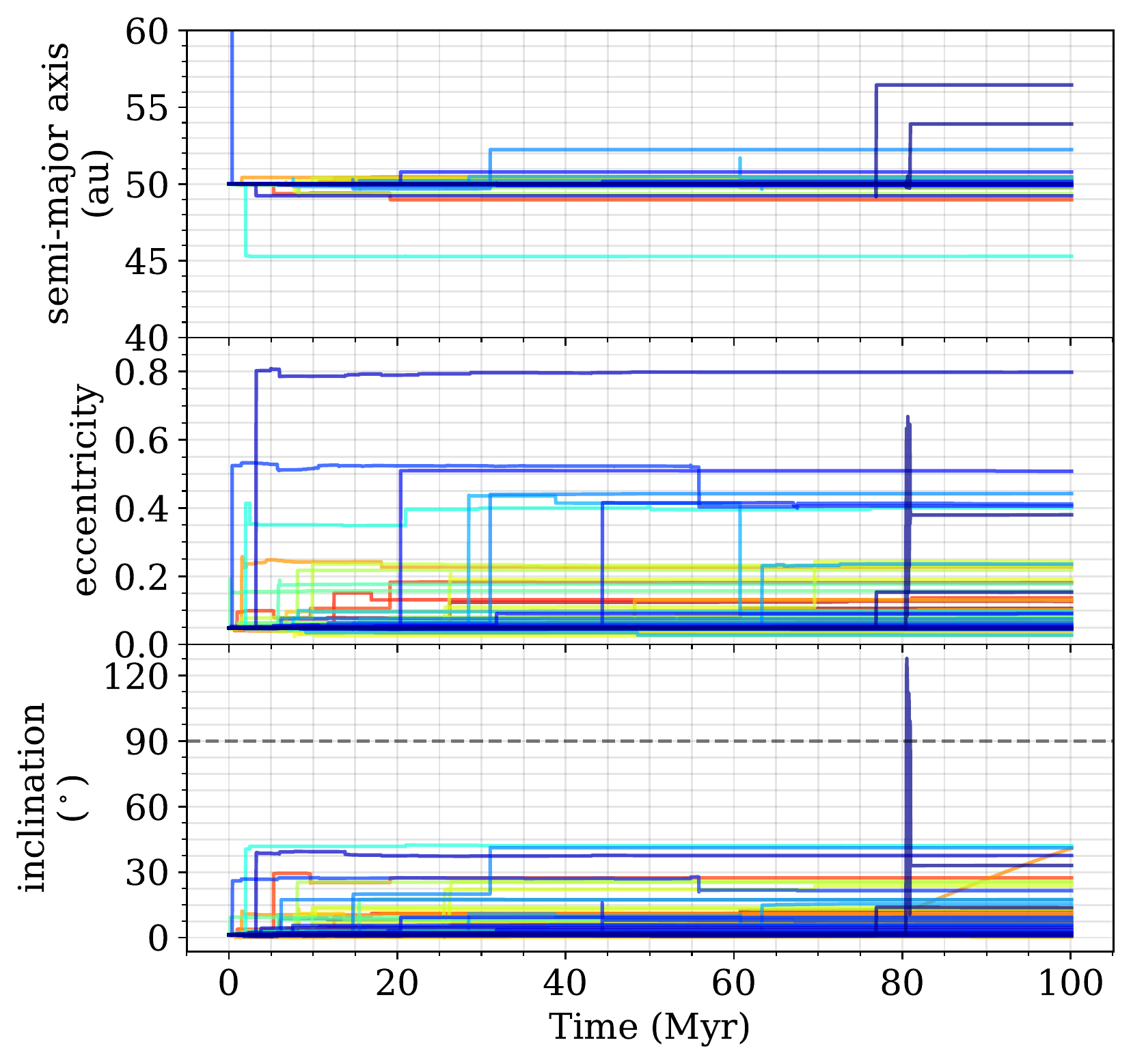}
    \caption{Evolution of the semi-major axis, eccentricity, and inclination of each planet in the ensemble of all planetary systems.}
    \label{fig.planet_aei_over_t}
\end{figure}

Encounters between planetary systems and neighboring stars may excite or eject the planet and may directly or indirectly affect the dynamics of the \pdps{} in the long term. Before examining the evolution of \pdps{}, we briefly analyze the evolution of the planetary orbits. Figure~\ref{fig.planet_aei_over_t} shows the evolution of the orbital elements of the ensemble of all planets.

Among the 100 planets, 16 escape from the planetary system within 100~Myr, 5 migrate inward (45.27, 48.98, 49.21, 49.22 and 49.37~au), 6 migrate outward (50.79, 51.70, 52.24, 53.91, 56.45 and 143.6~au) and the remaining planets retain their semi-major axes  $[49.5-50.5]$~au ($50~\mathrm{au}\pm 1\%$). Several planets acquire higher eccentricities, which allows their orbits to overlap with nearby \pdps{}. Except for one planet that suddenly obtains a high inclination ($>120^\circ$) at $t\approx 80$~Myr, all planets retain inclinations below $42.32^\circ$, which is slightly above Lidov-Kozai angle. For a more detailed analysis of the evolution of star-planet systems in star clusters, we refer to \citet{spurzem2009} and \citet{zheng2015}.

Since the masses of the planets are small compared to the stellar masses, the dynamical fate of a planet at $a_0 = 50$~au is almost identical to that of a \pdp{} with $a_0\approx 50$~au in a planetary system without planets. This will be discussed in Section~\ref{section:esc_fraction}, where we show that by the end of the simulation ($t=100$~Myr) the escape fractions of \pdps{} initially located at 45~au and 55~au are 16.0\% and 15.5\%, respectively. The latter values are in good agreement with the number of escaping planets (16\%). As we will discuss in Section~\ref{section:escaper}, 14 escaping planets are ejected from their planetary systems at speeds above the star cluster's local escape velocity. In other words, 86\% of the \fifju{} ultimately escape from the star cluster. This suggests that in young star clusters, a significant fraction of planets on relatively wide (50~au) orbits can remain bound to their host star during for 100~Myr.

Below, we will show that the contributions of the star cluster and the planet on planetary debris particles are not independent. The joint effect is complex, partly due to the interactions that arise from excitations of the planet and/or the \pdps{} following stellar perturbations. Excited planets tend to act as stirring rods, and alter the dynamics of the small particles.

\subsection{Fraction of escaping \pdps{}}
\label{section:esc_fraction}

In order to obtain statistically robust results, we focus on the combined properties of an evolving set of planetary systems in star clusters, rather than on individual systems. Here, and in Sections~\ref{section:survivor}-\ref{section:escaper}, the data shown is the combined result of the ensemble of all 100 planetary systems in the star cluster. The fastest evolution is experienced by the population of \pdps{} that are initially located near the planet, and the population that is most distant from the star. We quantify the stability of \pdps{} using the escape fraction $\chi(t,a_0)$, which depends on the initial semi-major axis, $a_0$, and on time, $t$:
\begin{equation}
    \chi(t,a_0) = \frac{\mathrm{Number~of~escapers~initially~on~}a_0\mathrm{~at~time~}t}{\mathrm{Total~number~of~\pdps{}~initially~on~}a_0}
    \quad .
\end{equation}
We will first study the dependence of $\chi$ on $a_0$ and $t$ individually, and subsequently constrain the influence of the planet by comparing $\chi(t,a_0)$ in simulations with and without a planet.

\subsubsection{Escape fraction: dependence on initial semi-major axis}
\label{sec.esc_fracion_over_a0}

\begin{figure}
    \begin{tabular}{l}
        \includegraphics[width=\columnwidth]{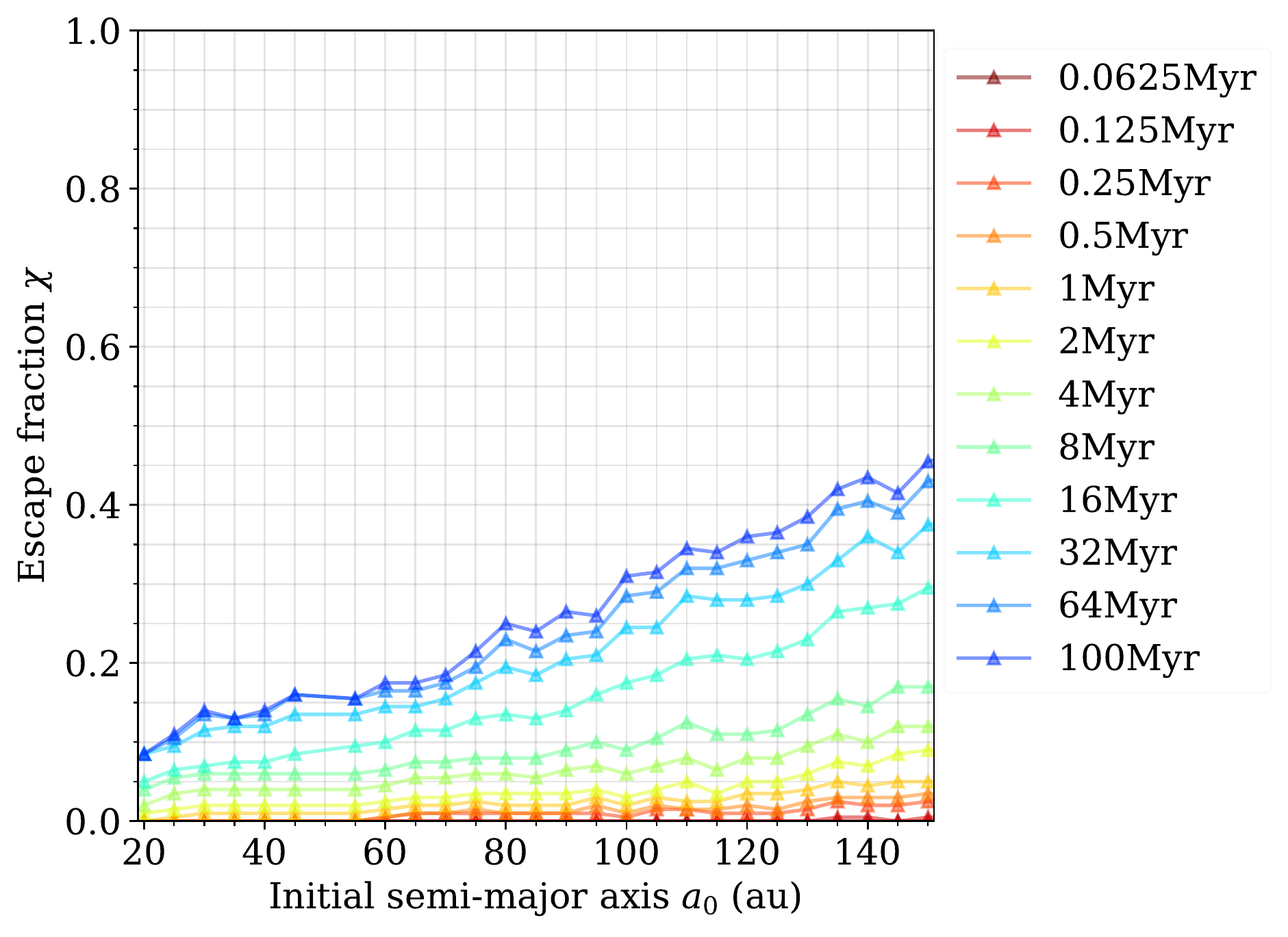} \\
        \includegraphics[width=\columnwidth]{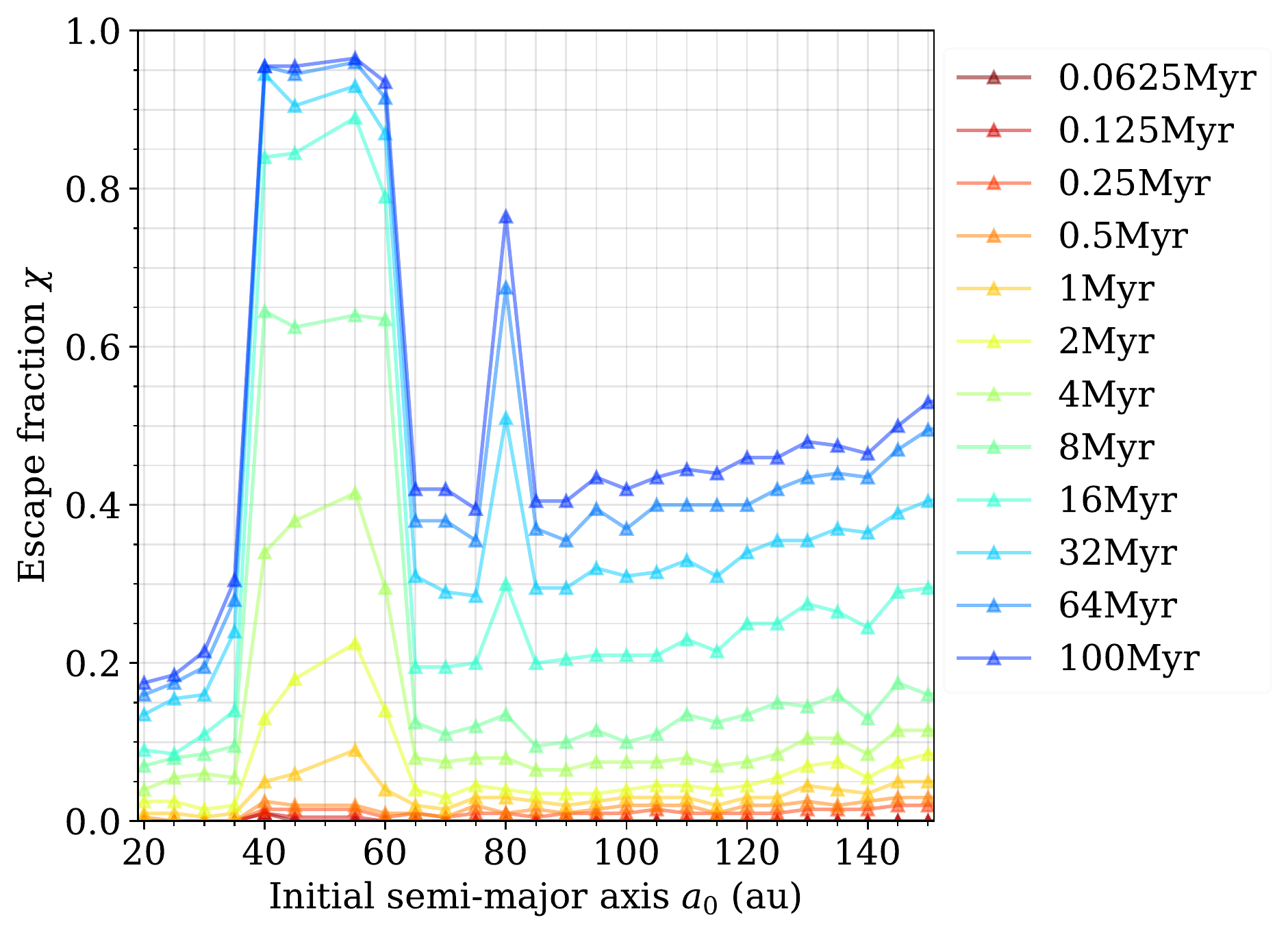} 
    \end{tabular}
    \caption{Escape fraction of \pdps{} as a function of initial semi-major axis, at different times, \ensemble{}. \emph{Top}: without planet. \emph{Bottom}: with planet. This figure shows results for $a_0 \le 150$~au (interior region).} \label{fig.8k_esc_fraction_over_a0_by_t_x200} 
\end{figure}

\begin{figure}
    \begin{tabular}{c}
        \includegraphics[width=\columnwidth]{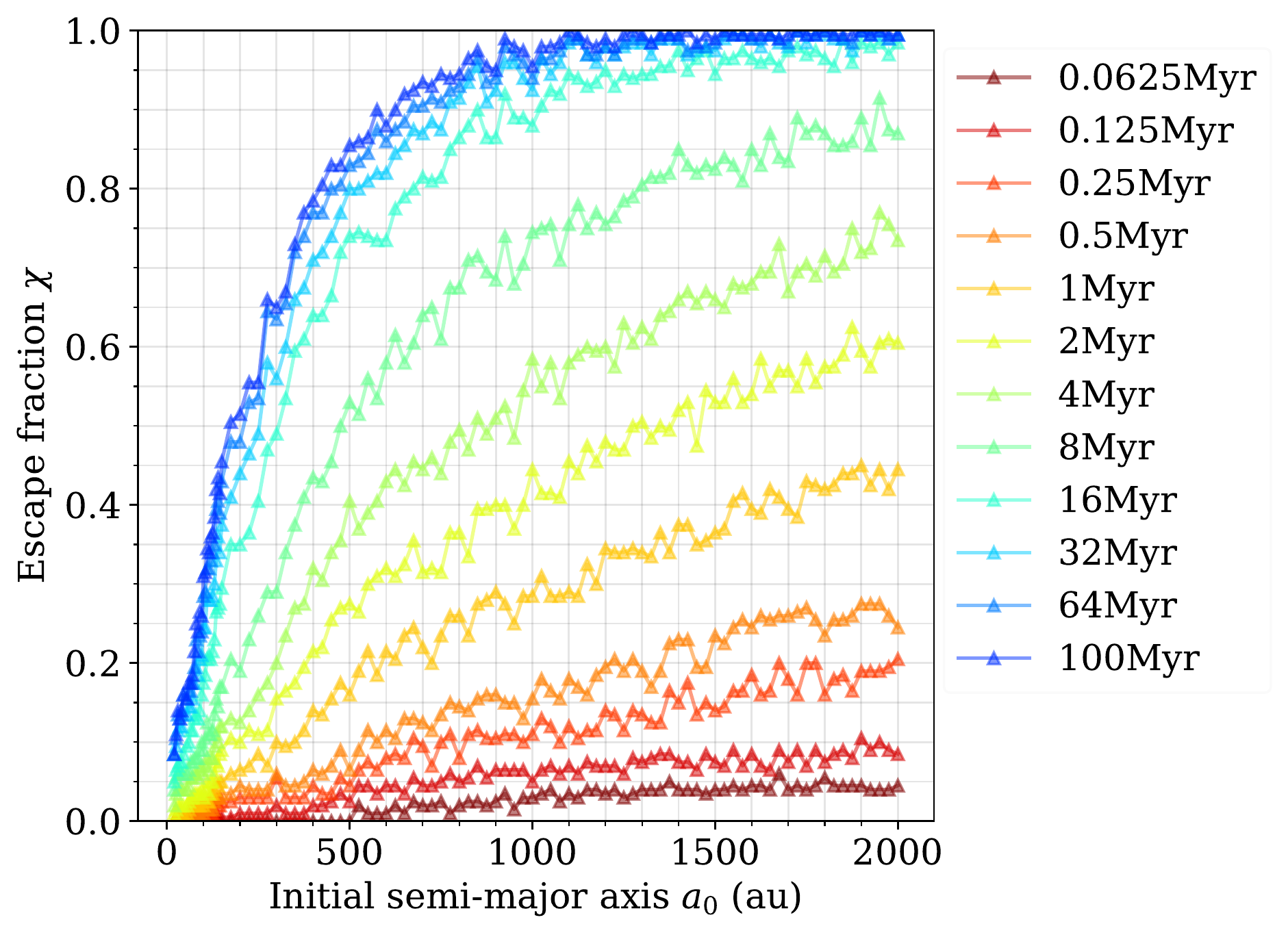} \\
        \includegraphics[width=\columnwidth]{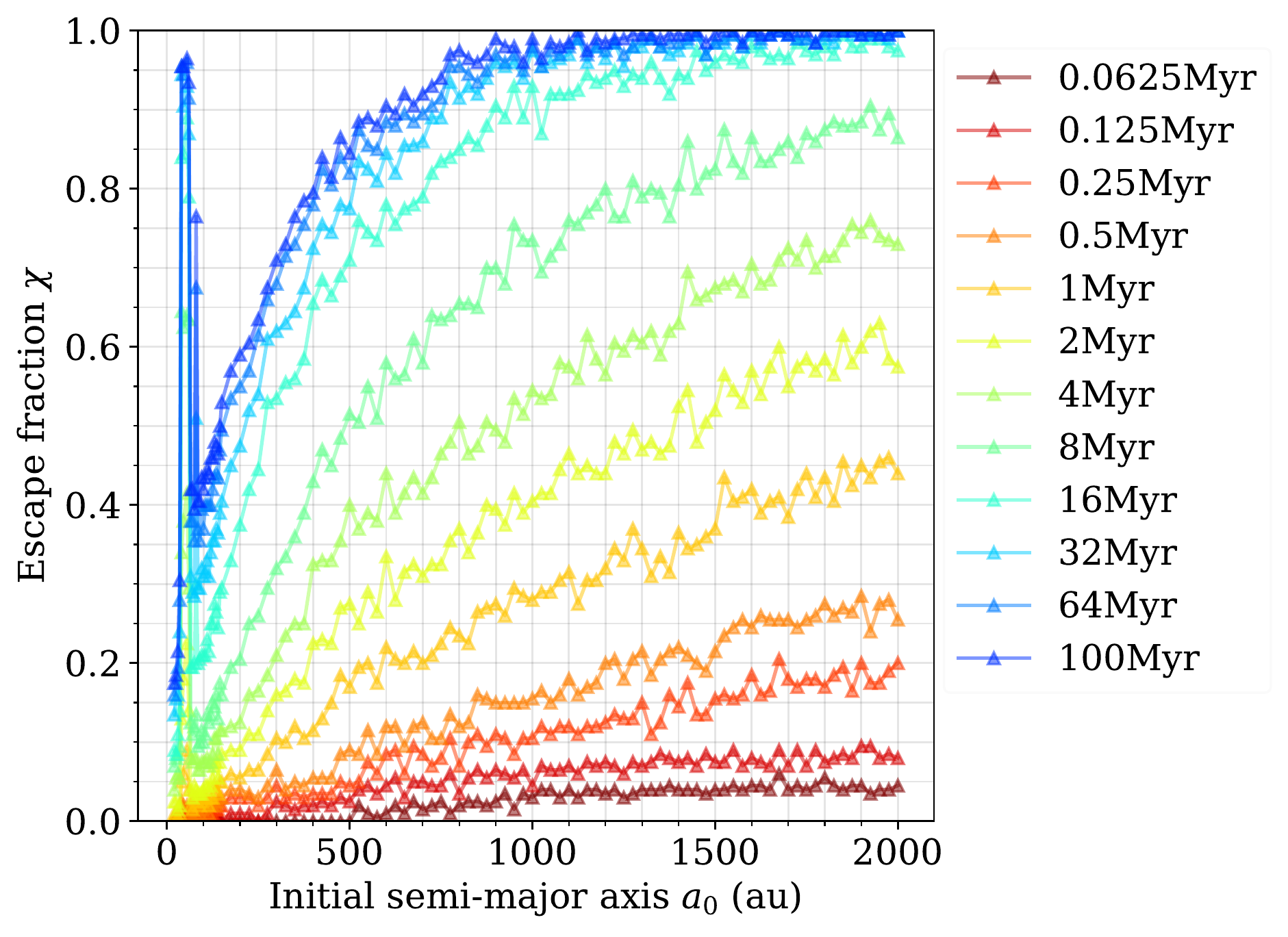} 
    \end{tabular}
    \caption{Same as in Figure~\ref{fig.8k_esc_fraction_over_a0_by_t_x200}, but now for the entire semi-major axis range.}
    \label{fig.8k_esc_fraction_over_a0_by_t}
\end{figure}

The escape fraction $\chi$ of \pdps{} as a function initial semi-major axis is shown in Figures~\ref{fig.8k_esc_fraction_over_a0_by_t_x200} and~\ref{fig.8k_esc_fraction_over_a0_by_t} for different times.  The dynamics of \pdps{} in the interior regions (Figure~\ref{fig.8k_esc_fraction_over_a0_by_t_x200}) of planetary systems is dominated by their host stars (and the planet, if present), although close encounters with neighboring stars can also affect this population under certain conditions. Most \pdps{} in the outer regions of the planetary systems are expelled from their planetary systems within the first tens of millions of years, as shown in Figure~\ref{fig.8k_esc_fraction_over_a0_by_t}, where the 16~Myr and 32~Myr curves exceed $\chi=0.85$. This trend extends inwards, down to a distance of roughly 1000~au. 

The behavior of \pdps{} with $a_0<1000$~au in the absence of a planet is shown in the top panels of  Figures~\ref{fig.8k_esc_fraction_over_a0_by_t_x200} and~\ref{fig.8k_esc_fraction_over_a0_by_t}. The innermost particles (i.e., \pdps{} with $a_0<70$~au) do not show a significant increase in the escape fraction as time increases, while the escape fraction of the intermediate-distance group of particles ($70<a_0/\mathrm{au}<1000$) continues to increase until throughout the simulation. The escape rate decreases beyond $t\approx 16$~Myr. At $t=64$~Myr, the escape rate drops to nearly zero.

The presence of the planet affects mainly \pdps{} with $a_0 \lesssim 300$~au. Results for models with the \fifju{} are shown in the bottom panels of Figures~\ref{fig.8k_esc_fraction_over_a0_by_t_x200} and~\ref{fig.8k_esc_fraction_over_a0_by_t}. The particles near the planet ($40-60$~au) show a notable increase in the escape rate compared to the model without a planet, and nearly reach unity by $t=100$~Myr. Comparing these results with the simulation of the isolated planetary system (Figure~\ref{fig.iso_esc_fraction_over_a0_by_t_x200}) indicates that the escape fractions for particles with $a_0 \le 150$~au (other than those at $40-60$~au and 80~au; see below) are not the effect of the planet alone, but are instead the joint effect of the \fifju{} and the perturbing stars. A prominent peak at $a \approx 80$~au is present in the bottom panels of Figures~\ref{fig.8k_esc_fraction_over_a0_by_t_x200} and~\ref{fig.8k_esc_fraction_over_a0_by_t}. This is caused by the 1:2 mean-motion resonant orbit at 79.37~au (for a planet at $a=50$~au). \Pdps{} in and near this region experience a high escape fraction. The escape fraction is also elevated for other particles initially within 150~au, and this can be attributed to the presence of the planet. A quantitative analysis of the influence is presented in Section~\ref{section:deltachi}.

The dependence of $\chi$ on $a_0$ in our study differs from those presented in \citet{cai2019}, who describe $f_\mathrm{surv}=1-\chi$ as a function of $\log_{10}(a_0)$ within 400~au. This hints that the star cluster properties may affect the functional relationship. \citet{veras2020} studied 2000 test particles initially on $40-1000$~au in each of the 11 planetary systems in a star cluster of 2000 stars. In their model, all test particles initially within 150~au are well protected from stellar perturbations in most of the 11 planetary systems. They find an overall escape fraction at $t=100$~Myr is 12.0\% for particles initially in the range $40-150$~au, and 37.0\% for particles in the range $40-1000$~au. To compare their data with our simulation, we use third-order spline interpolation to obtain escape fractions of test particles initially at their initial conditions (2000 particles uniformly distributed in the range $40-1000$~au), and our model gives 28.9\% for $a_0$ in $40-150$~au and 76.6\% for $a_0$ in $40-1000$~au. Although the escape statistics may depend weakly on the escape criteria, this suggests that the local stellar density around plays a crucial role in the stability of \pdp{} orbits, which we will investigate in a follow-up study.

\subsubsection{Escape fraction: dependence on time}

\begin{figure}
    \begin{tabular}{c}
        \includegraphics[width=\columnwidth]{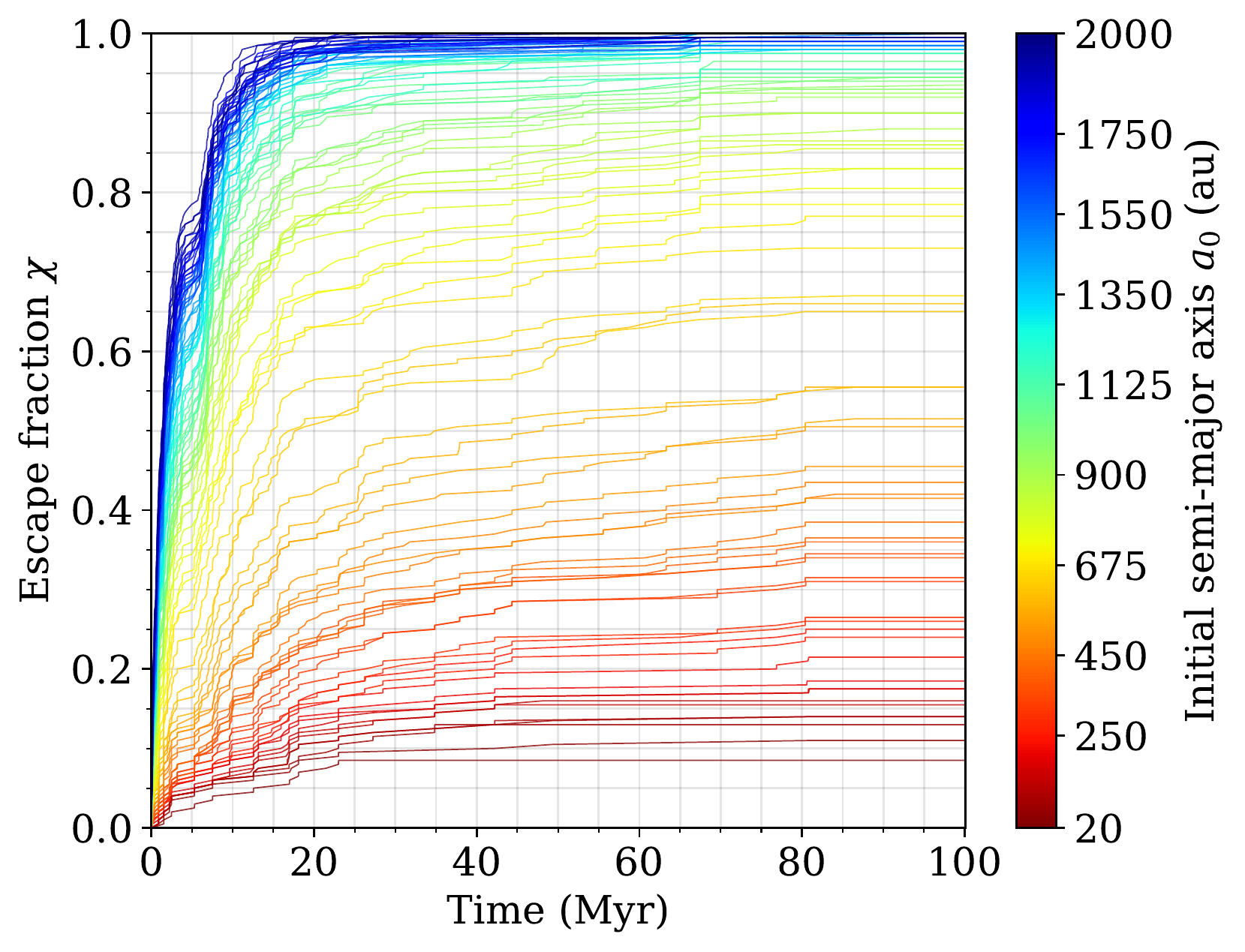} \\
        \includegraphics[width=\columnwidth]{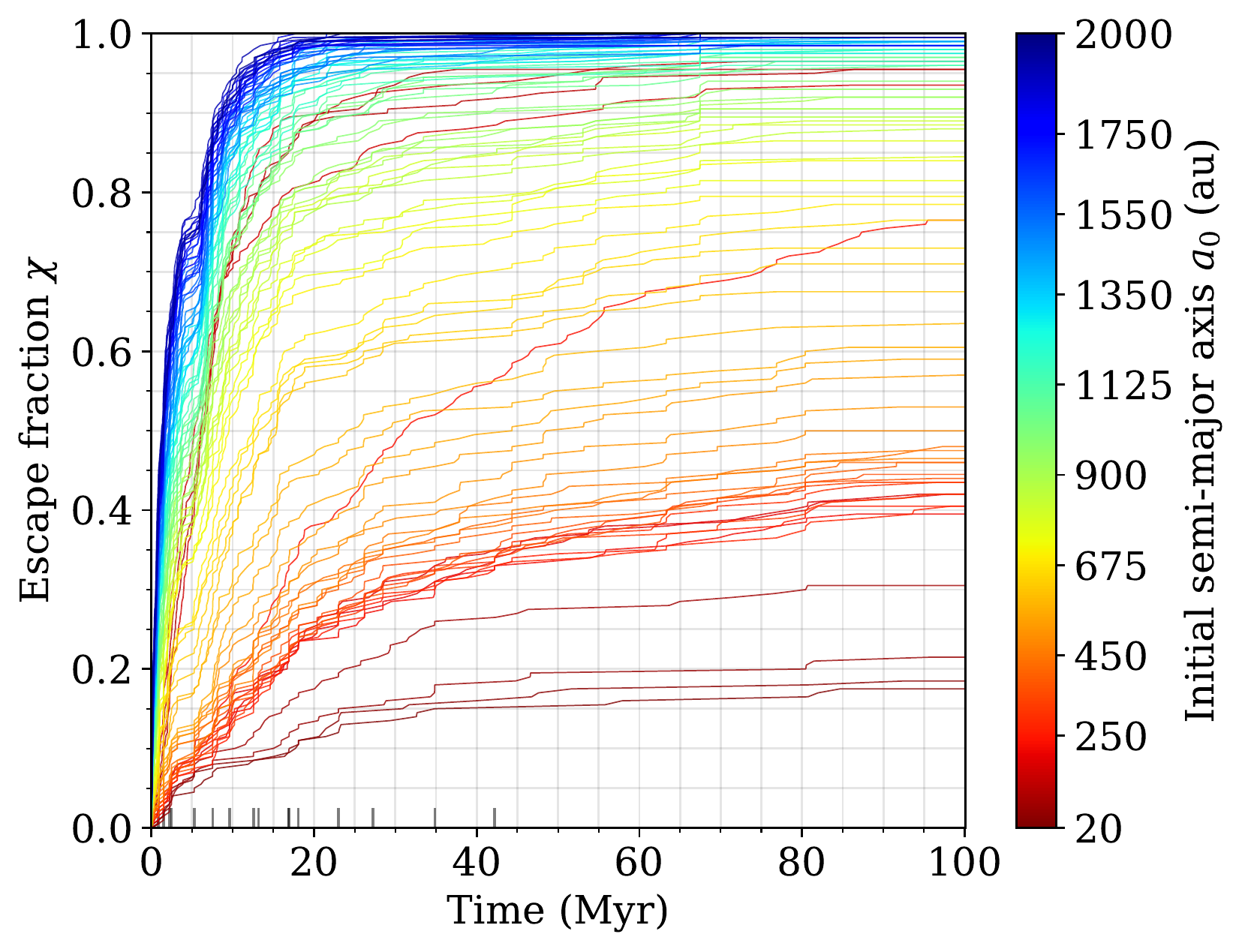} \\
    \end{tabular}
    \caption{Escape fraction as a function of time, \ensemble{}. \emph{Top}: without planet. \emph{Bottom}: with planet. Colors represent different initial semi-major axes. The black ticks at the bottom show the times at which planets escape from their planetary systems.  }
    \label{fig.8k_esc_fraction_over_t_by_a0}
\end{figure}

\begin{figure}
    \includegraphics[width=\columnwidth]{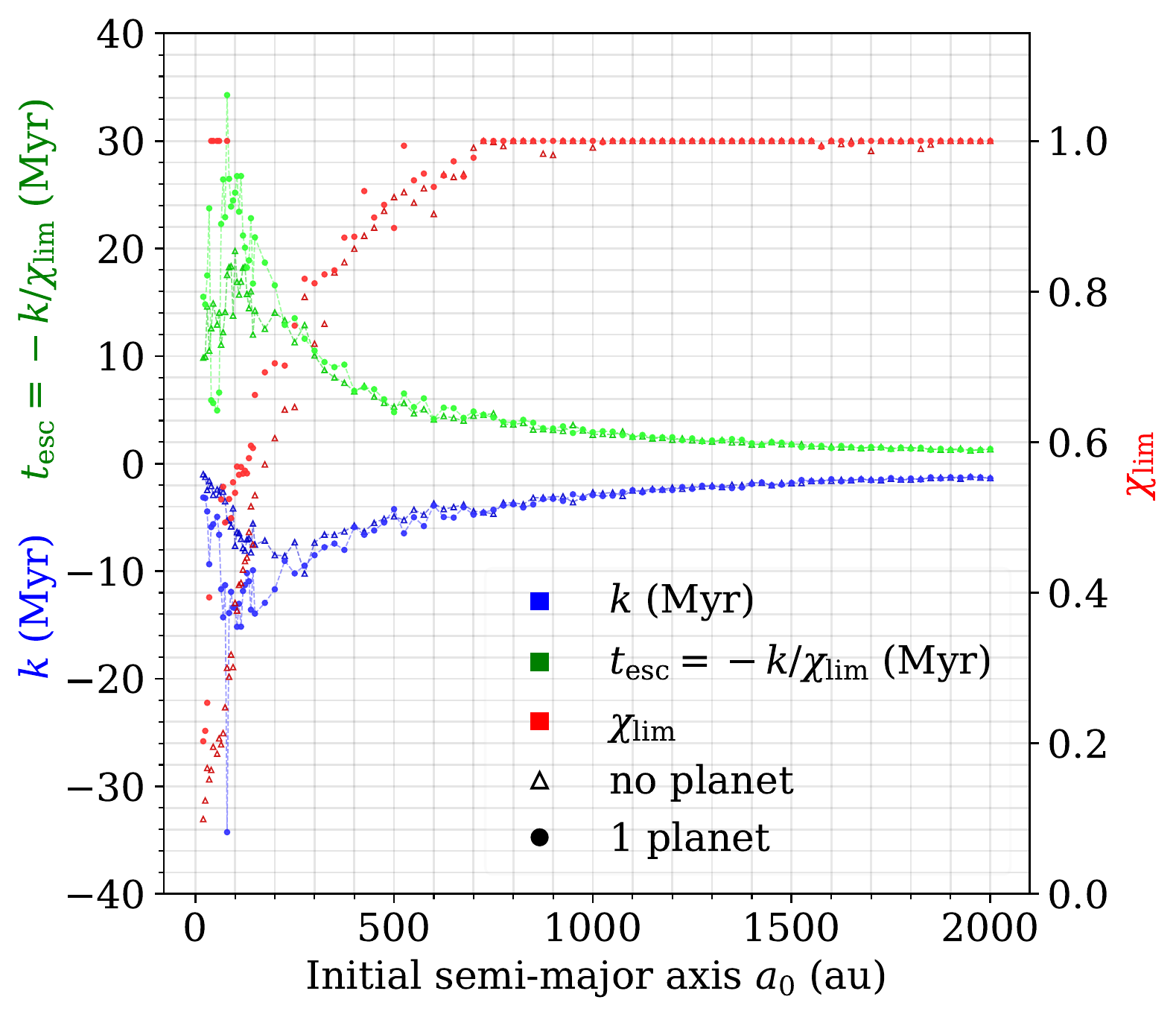}
    \caption{Values of the parameters $k$ (blue) and $\chi_\mathrm{lim}$ (red) of Equation~(\ref{eqn.fit}) and the characteristic escape timescale $t_\mathrm{esc}$ (green) as a function of initial semi-major axis. \emph{Triangles}: simulations without a planet; \emph{disks}: simulations with a planet.}
    \label{fig.8k_hyperbola_fit}
\end{figure}

Figure~\ref{fig.8k_esc_fraction_over_t_by_a0} shows the escape fraction $\chi$ as a function of time, $t$, for different initial semi-major axes, $a_0$. As all curves have shapes resembling hyperbolas, we model $\chi(t,a_0)$ as such;
\begin{equation}
    \chi(t, a_0)- \chi_\mathrm{lim}(a_0) = \frac{k(a_0)}{t - t_\mathrm{lim}}
    ~\quad .
    \label{eq:originalchi}
\end{equation}
Here, the horizontal asymptote is $\lim_{t \to \infty} \chi(t, a_0)= \chi_\mathrm{lim}(a_0)$, indicating the escape fraction when the evolution time tends to infinity. The quantity $k(a_0)$ determines the shape of the curve and thus $\chi$'s growth speed, where a large $|k|$ indicates fast growth. The parameter $t_\mathrm{lim}$ is the position of the vertical asymptote, which can be eliminated by the initial condition $\chi(t=0,a_0)=0$. 
This allows us to rewrite Equation~(\ref{eq:originalchi}) as
\begin{equation}
    \label{eqn.fit}
    \begin{split}
        \chi(t, a_0) & = \frac{k(a_0)}{t - k(a_0)/\chi_\mathrm{lim}(a_0)} + \chi_\mathrm{lim}(a_0) \\
                     & = \left(\frac{t}{k(a_0)} - \frac{1}{\chi_\mathrm{lim}(a_0)}\right)^{-1} + \chi_\mathrm{lim}(a_0)
                     ~\quad .
    \end{split}
\end{equation}

To estimate $k(a_0)$ and $\chi_\mathrm{lim}(a_0)$, we fit Equation~(\ref{eqn.fit}) to the data in Figure~\ref{fig.8k_esc_fraction_over_t_by_a0}. Figure~\ref{fig.8k_hyperbola_fit} shows the fitted parameters and escape timescales. The coefficients of determination ($R^2$) of the fits are all above 0.934, with a median value of 0.988, indicating that the escape fraction can be accurately described with Equation~(\ref{eqn.fit}). In general, $\chi_\mathrm{lim}(a_0)$ increases with $a_0$, reaching a maximum of 1.0 around 700~au. In Figure~\ref{fig.8k_esc_fraction_over_t_by_a0} we see that the green to blue curves ($a_0>1000$~au) eventually reach an escape ratio of 100\%, while the particles with $a_0=700-1000$~au particles (orange to green curves) do not. However, the fit suggests that almost all particles with $a_0>700$~au will eventually escape from their host stars. The presence of the planet significantly affects the fitted parameters for particles with $a_0<300$~au, raising their $\chi_\mathrm{lim}(a_0)$ to a certain degree, especially for particles near the planetary orbits ($40-60$~au).

In order to obtain a physical interpretation of $k$, we define the characteristic escape time $t_\mathrm{esc}$,
\begin{equation}
    \label{eqn.tesc}
    t_\mathrm{esc} \equiv -\frac{k}{\chi_\mathrm{lim}(a_0)} 
    ~\quad ,
\end{equation}
at which particles at initial semi-major axis $a_0$ reach half their final escape fraction\footnote{Note that $t_\mathrm{esc}$ differs from the concept of \textit{half-life} in nuclear physics, with the only special case for particles initially located far from the host star, that have $\chi_\mathrm{lim} = 1$, and simply have $t_\mathrm{esc}=-k$, which is also their \textit{half-life}.}, i.e., $\chi(t=t_\mathrm{esc}) =\tfrac{1}{2}\chi_\mathrm{lim}$. The parameter $t_\mathrm{esc}$ indicates how rapidly the escape fraction converges, and therefore provides a measure for the speed of evolution. $t_\mathrm{esc}$ decreases as $a_0$ increases when $a_0$ is large, but increases with $a_0$ for $a_0<100$~au. The presence of the planet does not influence $t_\mathrm{esc}$ for $a_0>200$~au. The region $a_0<200$~au can be classified into three subregions: (i) $40-60$~au, where $t_\mathrm{esc}$ decreases by approximately a half, (ii) $a_0\approx 80$~au, the 2:1 resonance peak, where $t_\mathrm{esc}$ is roughly doubled, and (iii) other regions within 200~au, where the contribution of the planet to $t_\mathrm{esc}$ depends on $a_0$. Except for the particles within the 10~au vicinity of the planet, which are rapidly expelled, the final escape fraction of particles within 300~au increases, while the time required to reach equilibrium is also longer. It is also interesting to note that the evolution of the escape fraction becomes almost identical to that of particles in the range of 65-150~au (except for 80~au) after adding the planet. This is also seen in the bottom panel of Figure~\ref{fig.8k_esc_fraction_over_a0_by_t_x200}. All particles initially in the distant region ($a_0>700$~au) are expected to escape within a relatively short time.

It is possible to perform additional fits of $k$ and $\chi_\mathrm{lim}$ as a function of $a_0$, to compactly describe the escape fraction as a multivariate function. However, since $k$ does not increase monotonically with $a_0$, with a turning point around $a_0\approx 300$~au, such a fit would be complex. This turning point is neither seen for the dense massive cluster of \citet[sec.~3.2]{cai2019}, nor in the sparse open clusters of \citet[sec.~3.3]{zheng2015}. We suspect that the dynamics in the region with $a_0<300$~au is different because the star cluster's influence cannot effectively reach this region. A better fit therefore requires inclusion of the star cluster parameters. To identify this relationship, we intend to further investigate the dependence on star cluster properties in a follow-up study.

\subsubsection{Difference in escape fractions}
\label{section:deltachi}

\begin{figure}
    \begin{tabular}{c}
        \includegraphics[width=\columnwidth]{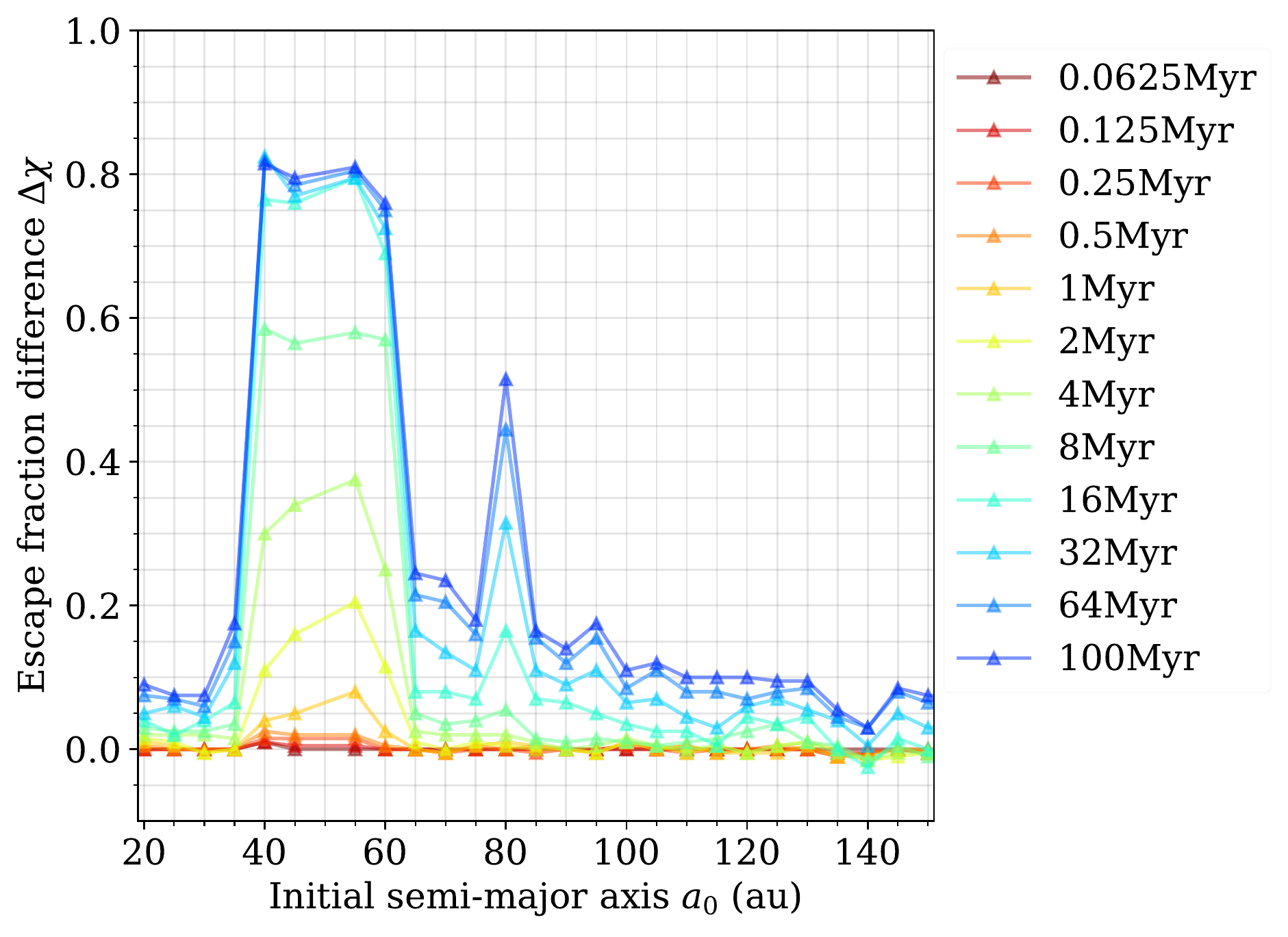} \\
        \includegraphics[width=\columnwidth]{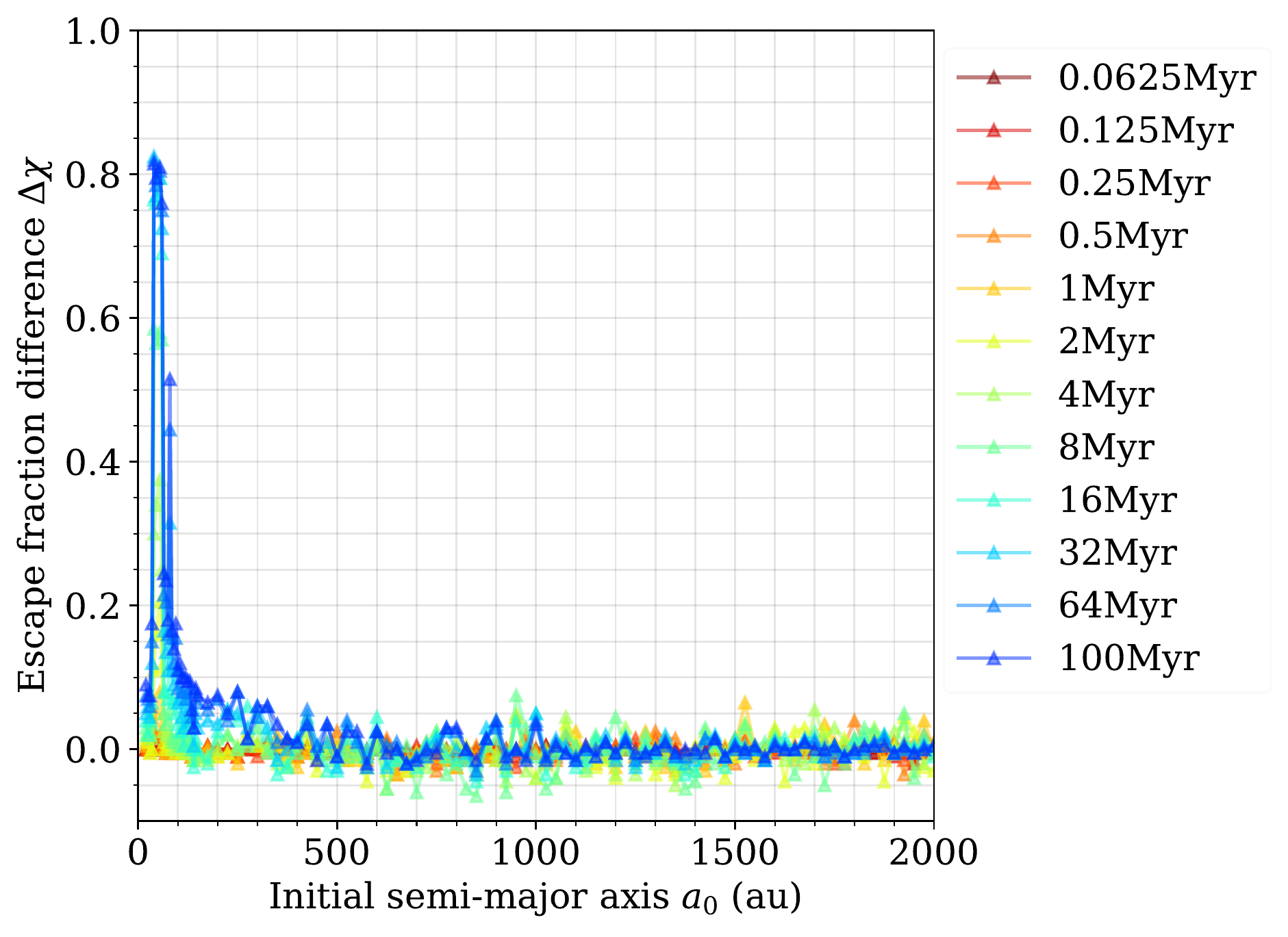} \\
    \end{tabular}
    \caption{Difference of the survival fraction, $\Delta\chi$, as a function of initial semi-major axis, \ensemble{}. {\em Top}:  $a_0 \le 150$~au (interior region). {\em Bottom}: full semi-major axis range.} 
    \label{fig.diff_Nesc_over_a0_by_t_8k}
\end{figure}

\begin{figure}
    \includegraphics[width=\columnwidth]{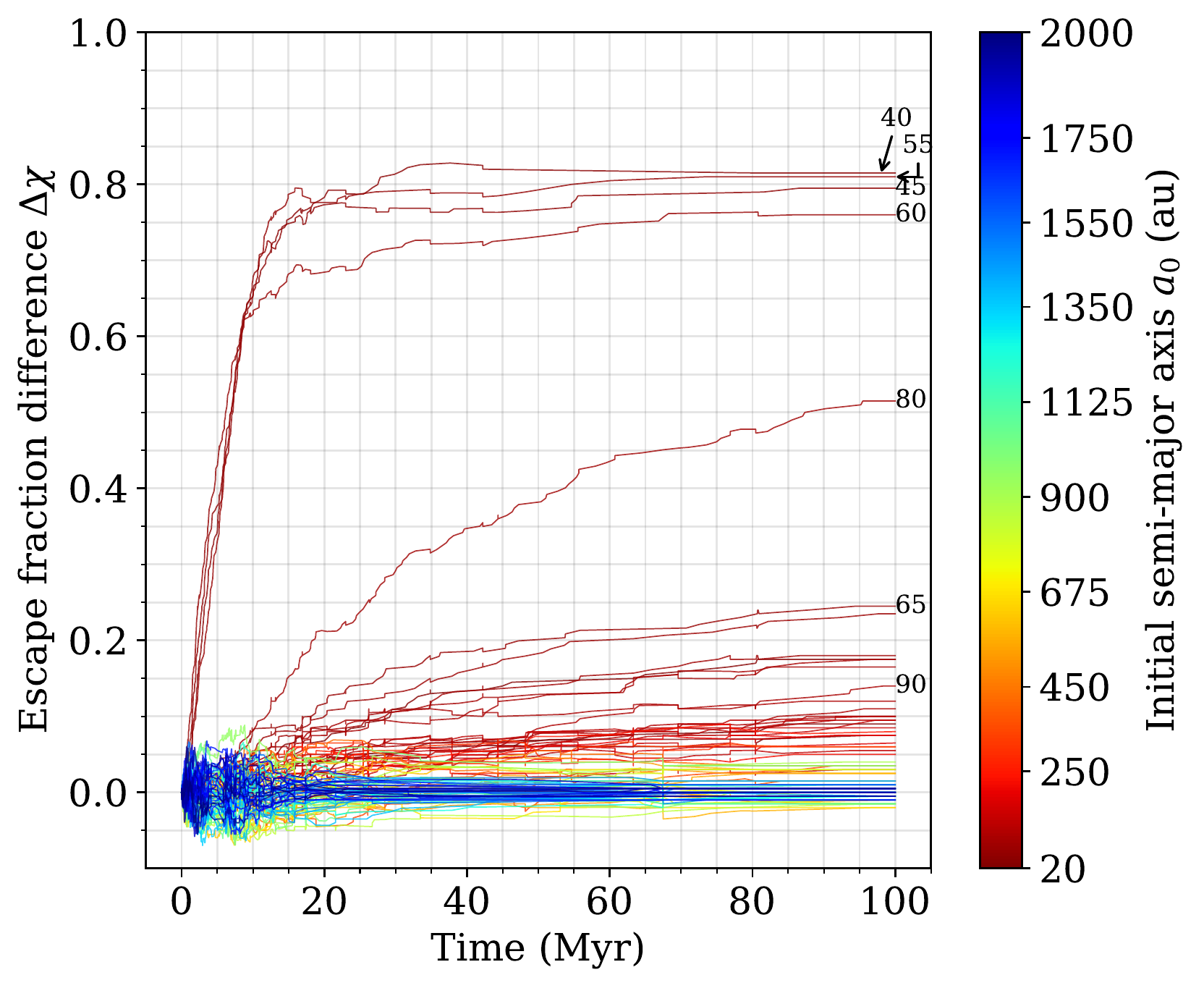} 
    \caption{Difference of survivor fraction, $\Delta\chi$, as a function of simulation time, \ensemble{}. Different colors indicate different initial semi-major axes. Values of $a_0$ for several curves with the highest $\Delta\chi$ are indicated.}
    \label{fig.diff_Nesc_over_t_by_a0_8k}
\end{figure}

We identify the influence of the planet on the escape fraction by comparing the escape fractions: 
\begin{equation}
    \Delta\chi(t,a_0) \equiv \chi_\mathrm{with-planet}(t,a_0) - \chi_\mathrm{without-planet}(t,a_0)
    \quad .
\end{equation}
We plot the difference in escape fractions as a function of initial semi-major axis and time in Figures~\ref{fig.diff_Nesc_over_a0_by_t_8k} and~\ref{fig.diff_Nesc_over_t_by_a0_8k}, respectively.

Both figures show that, generally, $\Delta\chi(a_0,t)>0$. This indicates that the presence of the planet is a destabilizing factor for the \pdps{} at most semi-major axes, and at nearly all times. Exceptions mostly occur in regions distant from the planet, which are indicated with the blue curves in Figure~\ref{fig.diff_Nesc_over_t_by_a0_8k}, during the first tens of Myr. Although some \pdps{} with large semi-major axes in simulations without planet have slightly higher escape fractions at the start, over longer times the difference tends to zero, as indicated by the blue curves in Figure~\ref{fig.diff_Nesc_over_t_by_a0_8k}, which oscillate around $\Delta\chi = 0$ with increasing $a_0$. 
Figure~\ref{fig.diff_Nesc_over_a0_by_t_8k} shows a noticeable dip at 140~au, where $\Delta\chi < 0$ when $t \le 16$~Myr. This may be caused by the relatively large statistical error. For example, $\Delta\chi(t=100\,\mathrm{Myr},a_0=40\,\mathrm{au}) \approx 0.03 \pm 0.10$. Particles at $a_0=140$~au are near the 14:3 mean motion resonance of the planet at 50~au, which is a high-order (i.e., weak) resonance, and is thus unlikely to play a stabilizing role during the early evolution of \pdps{} at $a_0 \approx 140$~au.

The escape fraction of the \pdps{} in the vicinity of the planet is most strongly affected, as indicated by the peaks in Figure~\ref{fig.diff_Nesc_over_a0_by_t_8k} around 50~au (40, 45, 55 and 60~au), with $\Delta\chi = 0.79_{-0.03}^{+0.02}$. This comes as no surprise, but the interesting peak at 80~au, the less obvious peak at 95~au, and the hint of a peak at 105~au indicate the effect of the mean motion resonances (2:1, 5:2 and 3:1 for the \fifju{}). Less prominent influences are seen for \pdps{} on other orbits with $a_0<130$~au, with the difference in escape ratio larger than 0.1, which experience a smooth growth over time, as shown in Figure~\ref{fig.diff_Nesc_over_t_by_a0_8k}.

The escape fraction of \pdps{} with $a_0>500$~au is barely influenced by the presence of a planet; the difference is less than 5\% at $t=100$~Myr, and the data fluctuate around zero (Figure~\ref{fig.diff_Nesc_over_a0_by_t_8k}).

The change in the escape rate difference over time ($\mathrm{d}\Delta\chi/\mathrm{d}t$) depends on the initial semi-major axis. For orbits near the planet, the difference increases remarkably fast within the first 10~Myr and then slows down at later times. The difference convergences at $t=100$~Myr, because most particles escape, while others migrate to safer orbits (see below and Section~\ref{section:survivor};  almost no particle remains in the neighborhood of the planet). In general, all curves monotonically increase, which indicates that the influence of the planet always accumulates. For all initial semi-major axes, $\mathrm{d}\Delta\chi/\mathrm{d}t$ is roughly constant before it reaches the limit (for example, in Figure~\ref{fig.diff_Nesc_over_t_by_a0_8k}, the four orbits near the planet during $0-10$~Myr and other orbits during $0-100$~Myr).

\subsubsection{Stability: classification of regions}
\label{section:classificationOfRegions}

\begin{figure}
    \includegraphics[width=\columnwidth]{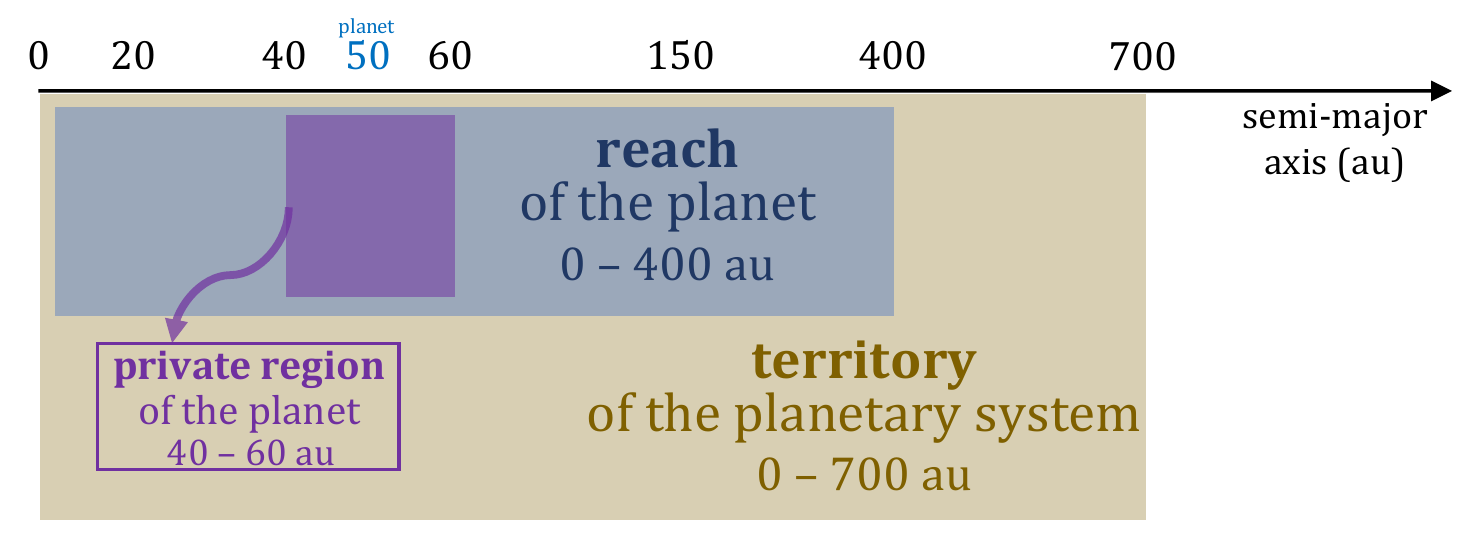} 
    \caption{Classification of regions. }
    \label{fig.division}
\end{figure}

Above we have shown that the stability of circumstellar \pdps{} in dense stellar environments depends on whether a planet is present, and on the properties of the neighboring stellar population. As shown in Figure~\ref{fig.division}, based on our findings, we classify the regions of our planetary system in the star cluster as follows:
\begin{enumerate}[leftmargin=0.6cm, labelsep=3pt, itemindent=-5pt]
    \item \textit{The private region of the planet} ($40-60$~au), where the \fifju{} clears all particles near its orbit.
    \item \textit{The reach of the planet} ($0-400$~au), within which the planet has a notable influence. Particles outside 400~au remain unaffected by the presence of the \fifju{}.
    \item \textit{The territory of the planetary system} ($0-700$~au), the region in which most particles remain part of the planetary system. Most particles outside this region are rapidly removed by stellar flybys. 
\end{enumerate}
Although the above results follow from the analysis of the escape fraction, we will show below that the classification is also useful in terms of the characteristics of both survivors and escapers. The location of the boundaries of the regions may vary, depending on the properties of the planet and the star cluster (see, e.g., \citet{cai2019}, where the border of the planetary system is 300~au in their young massive cluster). We aim to further investigate the classification of regions in the future.

\subsection{Properties of surviving particles}
\label{section:survivor}

\begin{figure*}
    \includegraphics[width=2\columnwidth]{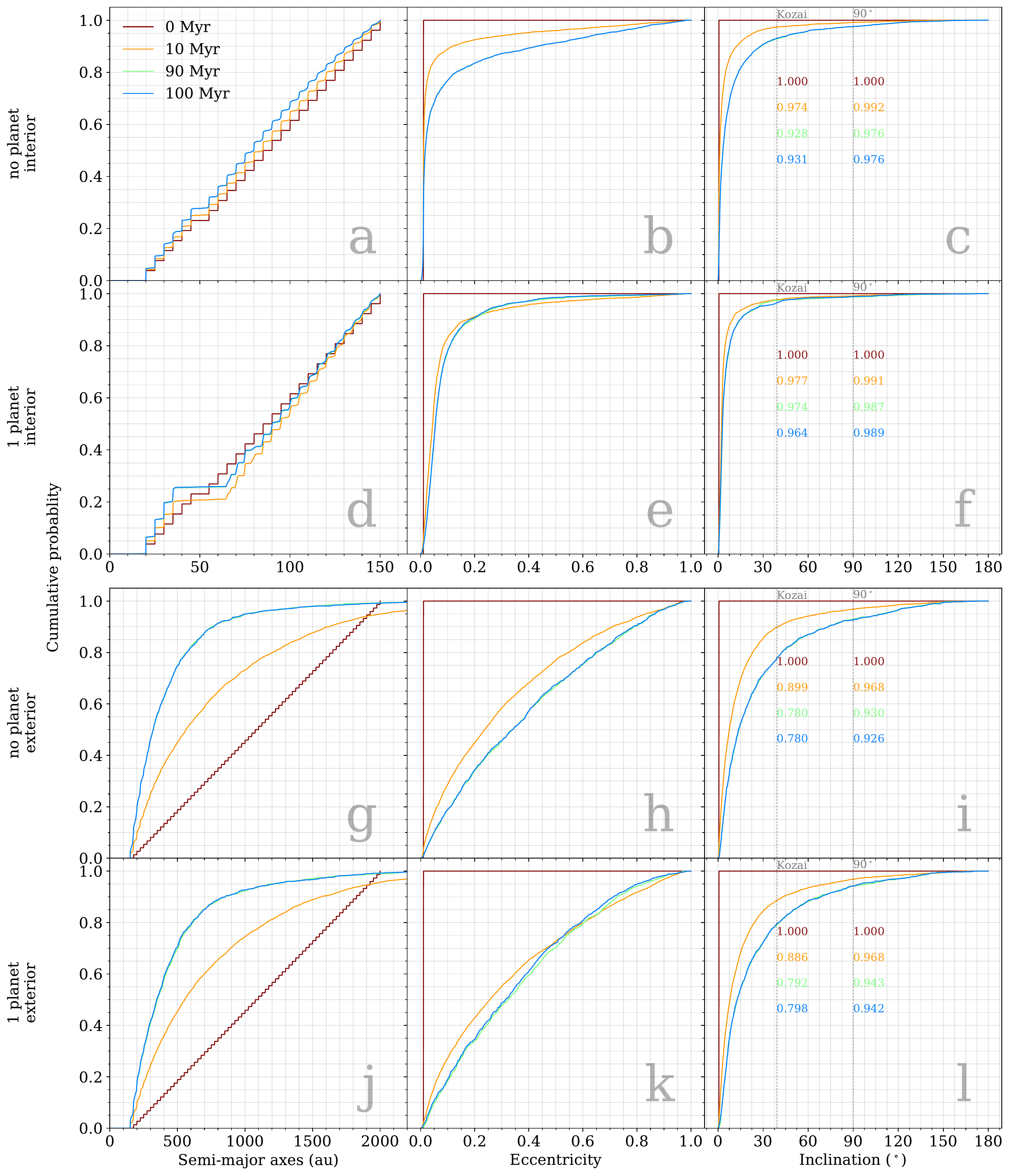} 
    \caption{Cumulative distribution functions of semi-major axis (left-hand panels), eccentricity (middle panels) and inclination (right-hand panels) at $t=0$, 10, 90 and 100~Myr of the \pdps{}, \ensemble{}. Results are shown for $a_0 \le 150$~au (top panels) and $a_0>150$~au (bottom panels). In the right-hand panels, the probabilities of particles with inclination within $39.2^\circ$ (labeled "Kozai") and with prograde orbits (labeled "$90^\circ$") are annotated for different simulation times. This figure was produced using the same data as in Figure~\ref{fig.8k_SURVIVORS_trans} but has been plotted differently to highlight the dependence on time.} 
    \label{fig.8k_SURVIVORS}
\end{figure*}

\begin{figure*}
    \includegraphics[width=2\columnwidth]{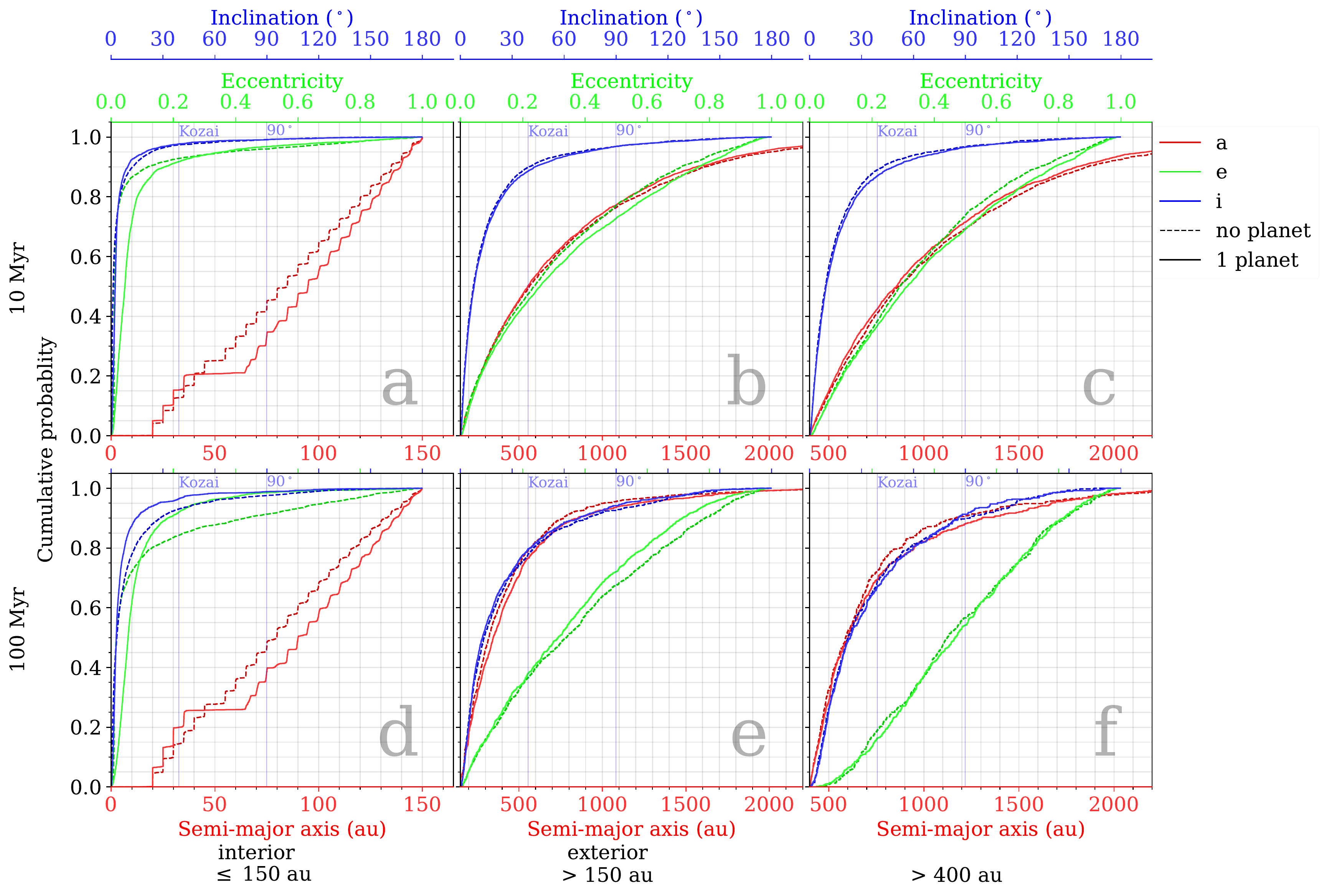} 
    \caption{Cumulative distribution functions of semi-major axis, eccentricity and inclination at $t=10$~Myr (top row) and $t=100$~Myr (bottom row) of \pdps{}, \ensemble{}. Results are shown for   $a \le 150$~au (left-hand panels), $a>150$~au (middle panels), and $a>400$~au (right-hand panels). The simulations with and without the planet are plotted with solid curves and dashed curves, respectively. Vertical lines in each panel are drawn at $i=39.2^\circ$ (the Kozai angle) and $i=90^\circ$. This figure was produced using the same data as in Figure~\ref{fig.8k_SURVIVORS}, but plotted differently to better compare the simulations with and without the planet.} 
    \label{fig.8k_SURVIVORS_trans}
\end{figure*}

\begin{figure*}
    \begin{tabular}{rl}
        \includegraphics[width=0.7\columnwidth]{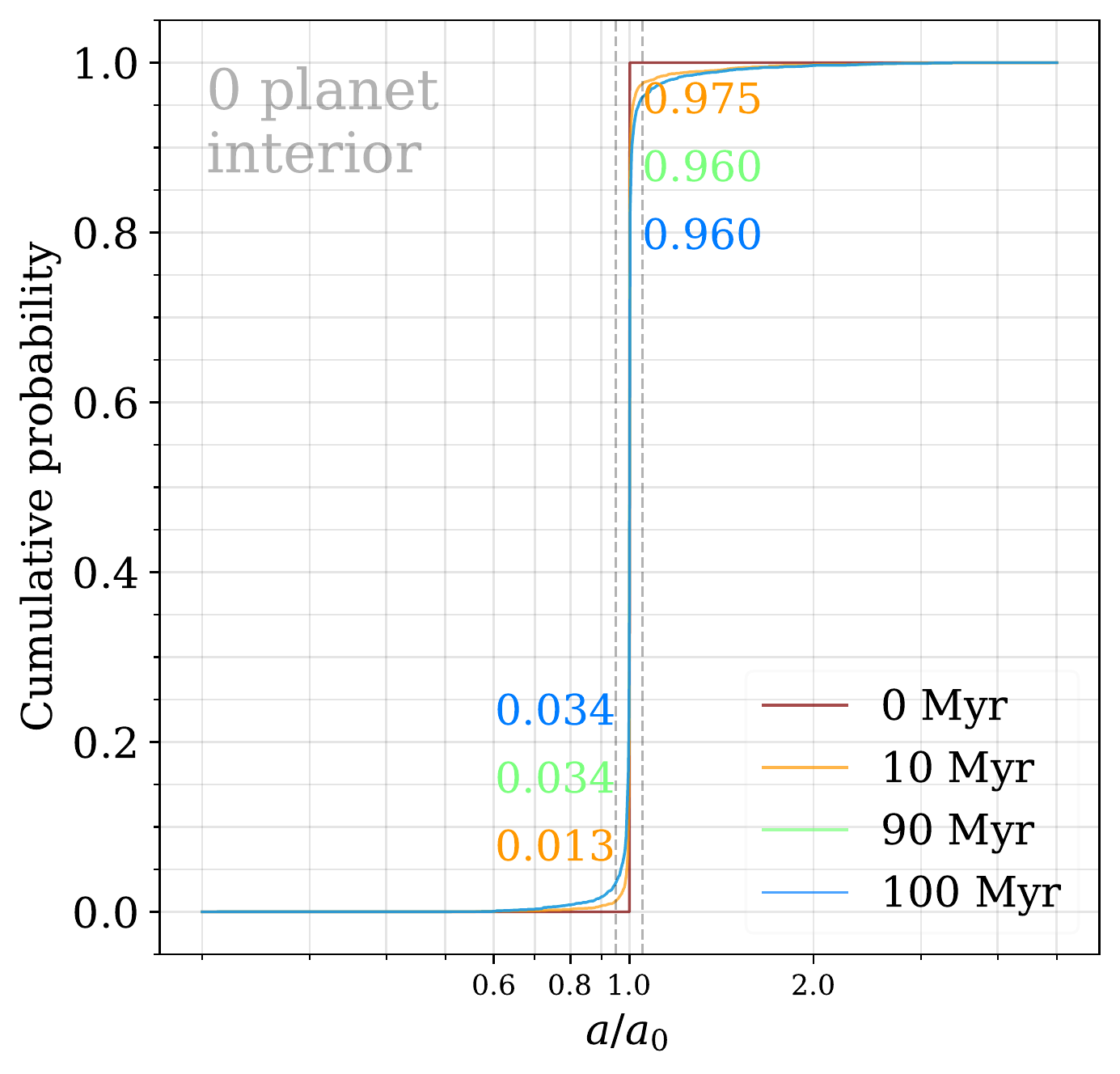} &
        \includegraphics[width=0.7\columnwidth]{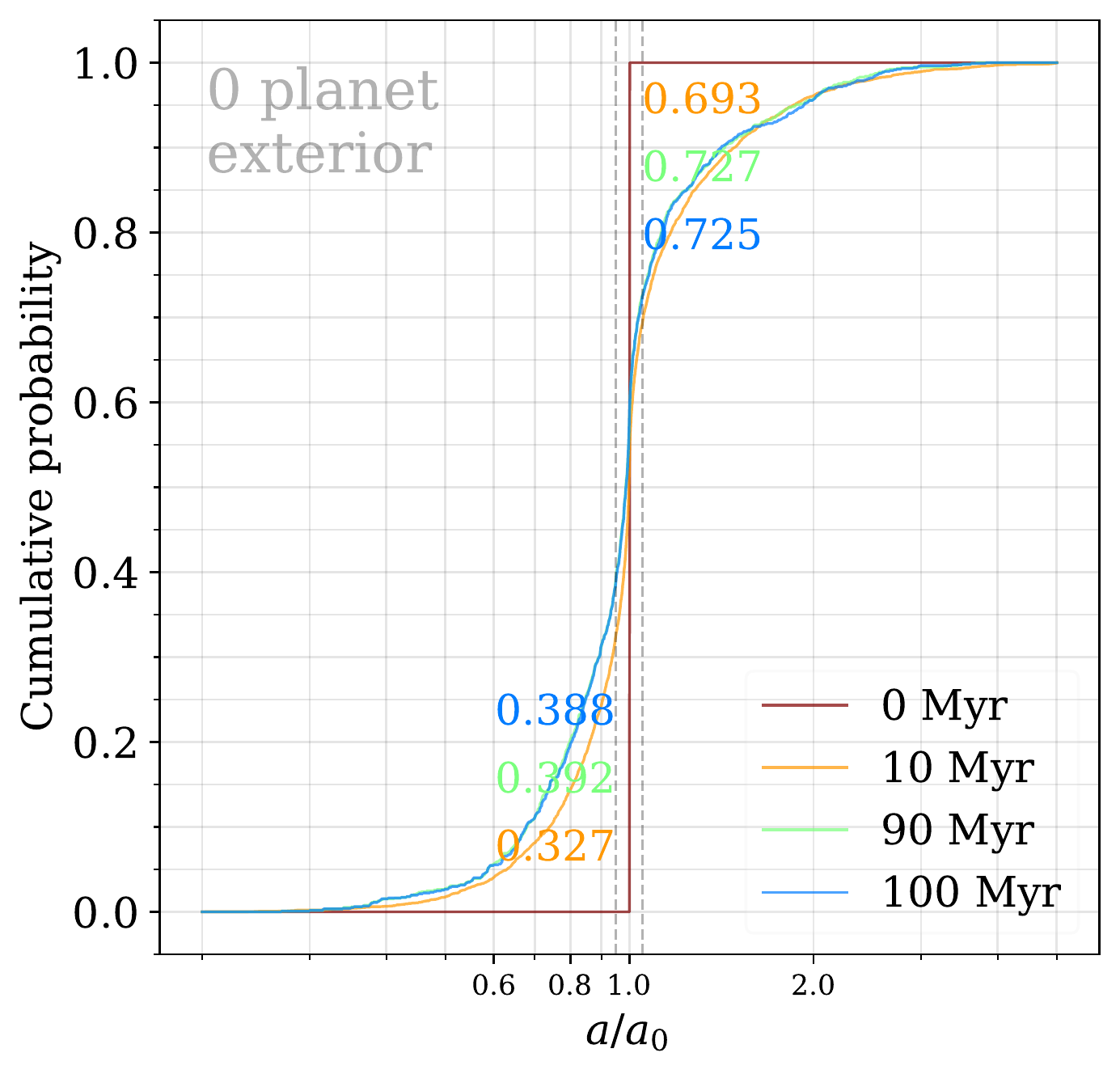} \\
        \includegraphics[width=0.7\columnwidth]{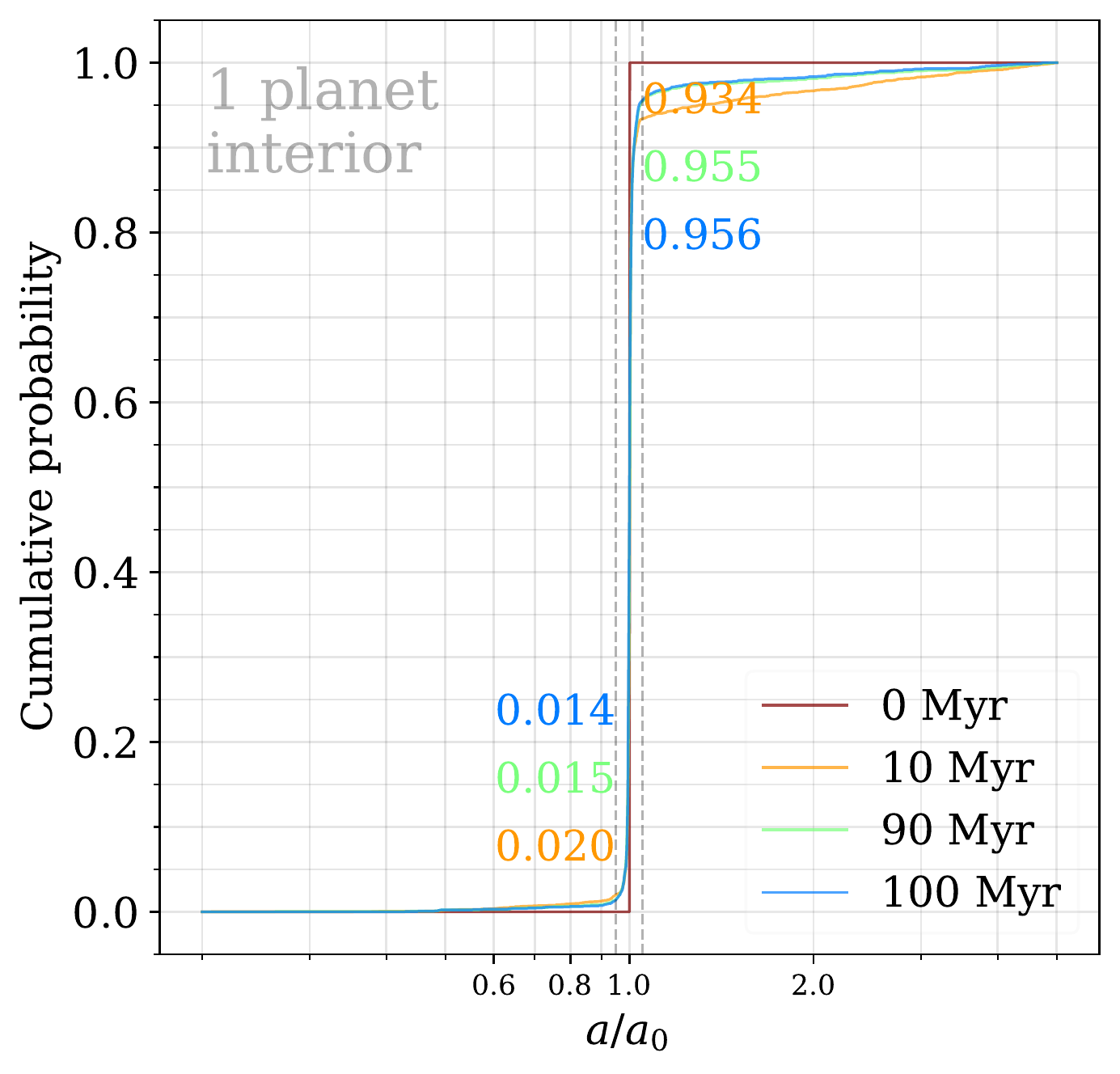} &
        \includegraphics[width=0.7\columnwidth]{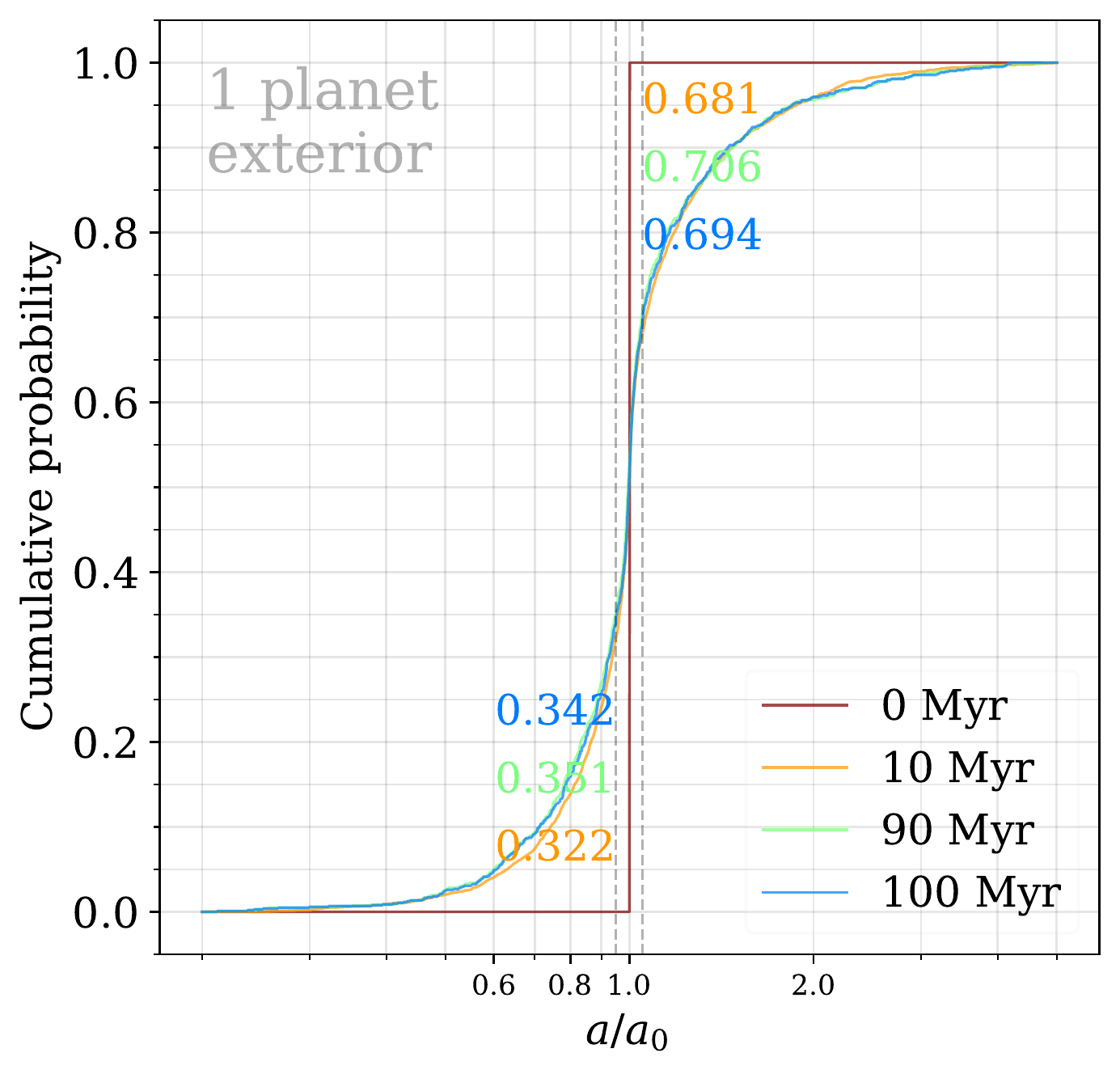} \\
    \end{tabular}
    \caption{Cumulative distribution of the ratio $a/a_0$, for all survivors, \ensemble{}. {\em Top}: without planet. {\em Bottom}: with planet. {\em Left}: $a_0 \le 150$~au. {\em Right}: $a_0>150$~au. The annotated numbers indicate the probability of $a/a_0=0.95$ and $1.05$ at different simulation times. The vertical dashed lines indicate $a/a_0=0.95$ and $a/a_0=1.05$. Note that we have only included surviving particles. Considering that many particles escape from the system (migration beyond 5000~au), the fraction of outward migrating particles will dominate if all particles are included.}
    \label{fig.8k_survivor_aovera0}
\end{figure*}

Figure~\ref{fig.8k_SURVIVORS} shows the evolution of the orbital parameters of the survivors (i.e., \pdps{} that remain bound to their host star). Following \citet{veras2020}, we plot the distributions of orbital parameters at 0, 10, 90, and 100 Myr.

For the interior of the planetary system ($a \le 150$~au), when the planet is absent, the proportion of particles close to the host star slightly increases over time (Figure~\ref{fig.8k_SURVIVORS}a), but the overall probability distribution does not change much. The proportion of particles with high eccentricities increases with time (Figure~\ref{fig.8k_SURVIVORS}b). Most particles, however, remain in near-circular orbits, with 50\% having $e<0.017$ and 75\% having $e<0.085$ at $t=100$~Myr. The presence of the planet reduces the fraction of particles near the initial orbit of the planet to almost zero within 10~Myr (Figure~\ref{fig.8k_SURVIVORS}d). Longer simulations lead to higher fractions of particles with small semi-major axes. The median eccentricity is $e=0.054$, with 75\% of the particles with $e<0.090$ and 90\% with $e<0.182$ (Figure~\ref{fig.8k_SURVIVORS}e). When the planet is included, the eccentricity evolution of interior \pdps{} is fast, as shown by the proximity of the distributions at $t=10$~Myr (orange curve) and $t=100$~Myr (blue curve) in Figure~\ref{fig.8k_SURVIVORS}(e).

For the exterior region ($a>150$~au), the fraction of particles with larger semi-major axes decreases dramatically over time. At $t=100$~Myr, 90\% of the particles have semi-major axes $a<768$~au in simulations where the planet is absent (Figure~\ref{fig.8k_SURVIVORS}g), and the eccentricity distribution becomes almost uniform (Figure~\ref{fig.8k_SURVIVORS}h). The influence of the planet is not significant in the exterior region, as the evolution of \pdps{} in this region is very similar for simulations with or without the planet (panels g vs j, and panels h vs k).

In the right-hand column of Figure~\ref{fig.8k_SURVIVORS}, we annotate for different times the probability of particles with inclination below the Kozai angle ($< 39.2^\circ$) and the probability of particles with prograde orbits. In general, more particles tend to gain higher eccentricity as time passes. At $t=100$~Myr, 2.4\% of interior surviving particles have retrograde orbits when no planet is present, and 1.1\% when the planet is present. In the exterior region, particles with high inclinations are more common than in the interior region. For example, 7.4\% obtain retrograde orbits without the planet, but when a planet is present, only 5.8\% are retrograde at $t=100$~Myr. The same is seen for particles with inclinations above the Kozai angle. When no planet is present, the fraction of particles is 6.9\% in the interior region and 22.0\% in the exterior region. When the planet is present, 3.6\% of the particles in the inner region and 20.2\% of the particles in the outer region obtain inclinations above the Kozai angle. This is caused by a combination of two processes: (i) the planet prevents the \pdps{} from obtaining higher inclinations during/after stellar close encounters, (ii) the planet expels particles with high inclinations. Future studies may shed further light on each of these contributions.

It is worth noticing that the curves for 90~Myr and 100~Myr mostly overlap in all panels of Figure~\ref{fig.8k_SURVIVORS}. This indicates that the entire system does not change significantly during the final 10~Myr of the simulation, and indicates that our choice for the integration time of 100~Myr is adequate.

To further compare the differences between the simulations with and without the planet, we present the results in a different form in Figure~\ref{fig.8k_SURVIVORS_trans}. For clarity, we only use the data for $t=10$~Myr and $t=100$~Myr. For the interior region (Figure~\ref{fig.8k_SURVIVORS_trans}a, d), the planet expels most of the particles in its neighborhood. In the simulation with the planet, the inclinations of surviving particles are smaller. As a consequence, fewer \pdps{} have inclinations larger than the Kozai angle or have retrograde orbits, as discussed above.

For eccentricities of interior particles, we compare the resulting eccentricity distributions (Figure~\ref{fig.8k_SURVIVORS_trans}) with those of the planetary system that evolved in isolation (Figure~\ref{fig.iso_CDF_e_in}). The isolated planetary system obtains a nearly uniform distribution in the range $0<e<0.1$, for more than 95\% of the \pdps{}. This trend is also seen in the simulations where the star cluster is modeled, and the planet is included. The solid green curve in Figure~\ref{fig.8k_SURVIVORS_trans}(d) suggests more than 80\% particles obtain $0\leq e < 0.1$, with a nearly uniform distribution in this range. This is not seen in the simulation without the planet.

We expect that the influence of the planet on the \pdps{} in the exterior particles is negligible at early times ($t<10$~Myr) (Figure~\ref{fig.8k_SURVIVORS_trans}b), and this is indeed observed. However, at 100~Myr (Figure~\ref{fig.8k_SURVIVORS_trans}e), the average eccentricity and inclination of the \pdps{} are smaller for simulations in which the planet is present. The semi-major axis distribution of the exterior particles in the simulations with and without the planet, on the other hand, are almost identical.

The right-hand column of Figure~\ref{fig.8k_SURVIVORS_trans} shows the orbital element distribution of \pdps{} beyond 400~au, where the presence of the planet has little effect on the \pdps{} in this region. This is consistent with our statement of region classification in Section~\ref{section:classificationOfRegions}. 

In almost all panels of Figure~\ref{fig.8k_SURVIVORS}, the orange curves (90~Myr) and blue curves (100~Myr) are closer when the planet is present, except when comparing panel (d) with panel (a). Thus, in general, the presence of the planet accelerates the evolution of the \pdps{} in the star cluster, for particles both near and far from the planet, as a consequence of scattering events and secular evolution resulting from perturbed orbits.

We also examine the migration of surviving \pdps{} in Figure~\ref{fig.8k_survivor_aovera0}. For simplicity, a particle is considered to have migrated inward if its semi-major axis decreases by 5\% or more from its initial value, and outward if it increases by 5\% or more. When a planet is added to the simulations, the inward migration of the surviving particles is reduced from 3.4\% to 1.4\% (at the end of the simulation) for the interior region, and from 38.8\% to 34.2\% for the exterior region. The presence of the planet also increases the fraction of surviving \pdps{} that migrate outward: the ratio increases from 4.0\% to 4.4\% for the interior region and increases from 27.5\% to 30.6\% for the exterior region. The planet thus influences the migration pattern of \pdps{}: more surviving particles experience outward migration, and fewer experience inward migration, when the planet is present.

\subsection{Properties of escapers} 
\label{section:escaper}

\begin{figure}
    \includegraphics[width=\columnwidth]{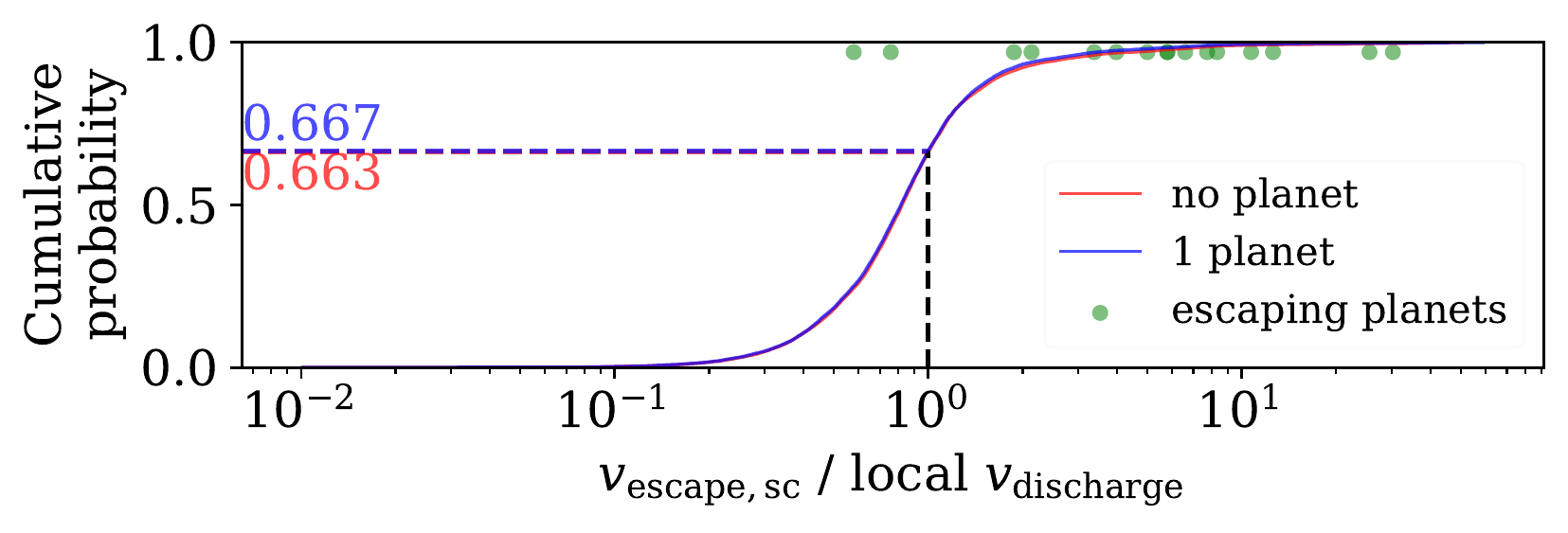}
    \caption[Caption for LOF]{Cumulative distribution of the ratio of the speed at which \pdps{} escape from their host systems ($v_\mathrm{escape,sc}$) to the local discharge barrier speed (local $v_\mathrm{discharge}$) in the star cluster, \ensemble{}. The corresponding values for the sixteen escaping planets are indicated with green dots. When $v_\mathrm{escape,sc}~/~\mathrm{local}~v_\mathrm{discharge} > 1$, the \pdp{} or planet is likely to immediately leave the star cluster. Otherwise, it escapes from the planetary system but remains gravitationally bound to the star cluster. Such object may be re-captured by another star, or may escape from the cluster at a later time.}
    \label{fig.8k_CDF_escapers_speed_over_vesc}
\end{figure}

\begin{figure*}
    \begin{tabular}{cc}
        \includegraphics[width=\columnwidth]{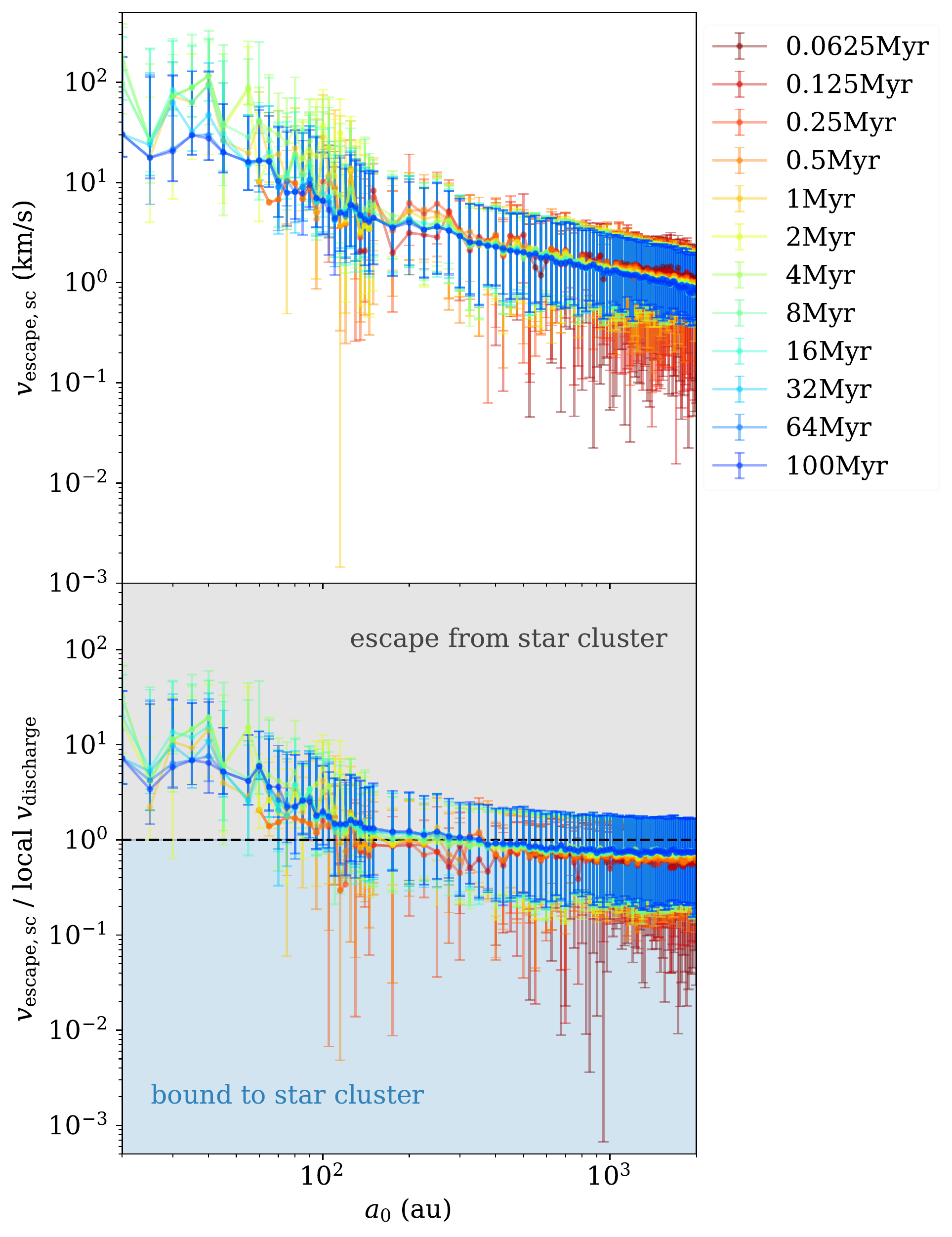} &
        \includegraphics[width=\columnwidth]{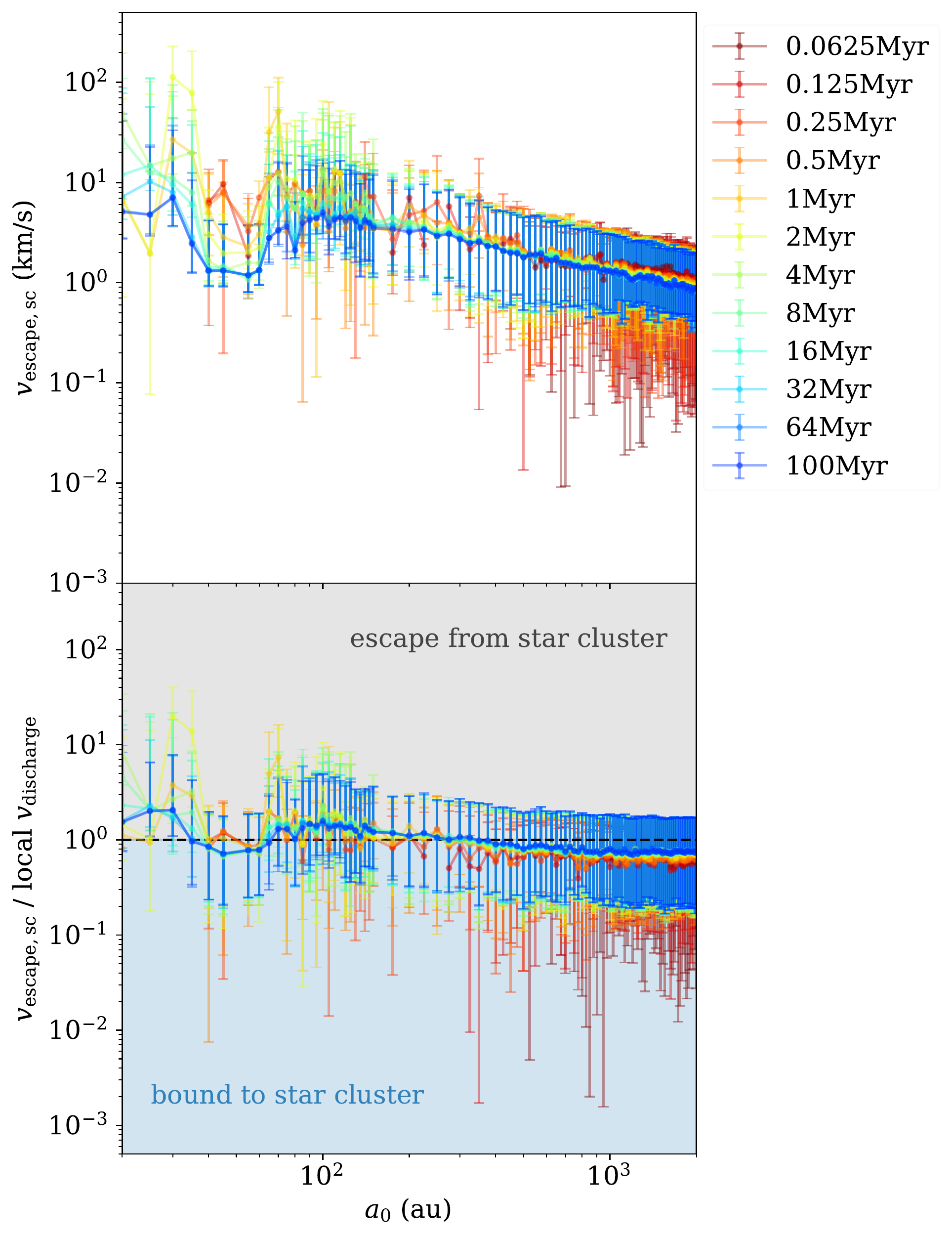} \\
    \end{tabular}
    \caption{Statistics of escapers, \ensemble{}. \emph{Left}: without planet. \emph{Right}: with planet. \emph{Top}: escape speed as a function of initial semi-major axis. \emph{Bottom}: ratio of escape speed ($v_\mathrm{escape,sc}$) to the local discharge barrier speed of the star cluster (local $v_\mathrm{discharge}$). The solid dot indicates the median, while the error bar represents the range between the first quartile (25\%) to the third quartile (75\%) of the data. Ranges in which \pdps{} escape from the star cluster are shaded gray, while ranges in which \pdps{} remain gravitationally bound to the star cluster are shaded blue.}
    \label{fig.8k_0_escapers_speed_over_a0all}
\end{figure*}

After a \pdp{} has been expelled from its host system, it becomes a free-floating particle in the star cluster. Depending on its speed and location, it may remain gravitationally bound to the cluster, or it may immediately escape. To avoid ambiguity, we use the word \textit{escape} to denote the departure of \pdps{} from a planetary system and the term \textit{discharge} to denote the departure of \pdps{} from a star cluster.

A \pdp{} is marked as having escaped from the planetary system when its distance to the host star exceeds the critical value (see Section~\ref{section:method}). Its corresponding velocity-at-infinity relative to the host star, $v_\infty$, is
\begin{equation}
    v_\infty = \sqrt{v_\mathrm{escape,pl}^2 - \frac{2G(M_*+M_\mathrm{planet})}{r}}
    ~\quad ,
\end{equation}
where $v_\mathrm{escape,pl}$ is the velocity relative to the host star when the particle is identified as having escaped, $M_*$ is the host star mass, and $M_\mathrm{planet}$ is the mass of the planet (if present). After a particle has escaped, it travels approximately in the direction of escape with a speed $v_\infty$. Its escape velocity vector $\vec{v}_\mathrm{escape,sc}$ relative to the \textbf{s}tar \textbf{c}luster center is
\begin{equation}
    \vec{v}_\mathrm{escape,sc} = \vec{v}_{\infty} + \vec{v}_\mathrm{host}
    ~\quad ,
    \label{eqn.vesc_sc}
\end{equation}
where $\vec{v}_{\infty}$ is the velocity vector in the escape direction with magnitude $v_{\infty}$, and $\vec{v}_\mathrm{host}$ is the velocity vector of the host star. A particle with mass $m$ directly escapes from the star cluster if it has positive total energy $E_\mathrm{total}$:
\begin{equation}
	E_\mathrm{total} = \tfrac{1}{2}mv_\mathrm{escape,sc}^2 + m\phi_\mathrm{particle} \ge 0
    \quad ,
\end{equation}
where $v_\mathrm{escape,sc} = |\vec{v}_\mathrm{escape,sc}|$, and $\phi_\mathrm{particle}$ is the gravitational potential at the location of the \pdp{} in the cluster. We define the discharge barrier speed $v_\mathrm{discharge}$, which is the $v_\mathrm{escape,sc}$ when $E_\mathrm{total} = 0$, as
\begin{equation}
    v_\mathrm{discharge} = v_\mathrm{escape,sc} (E_\mathrm{total} = 0) = \sqrt{-2\phi_\mathrm{particle}}
    ~\quad .
    \label{eq:discharging}
\end{equation}
The local gravitational potential at the location of the escaping particle is similar to that at the location of the host star. We thus adopt the approximation $\phi_\mathrm{particle} \approx \phi_*$, where $\phi_*$ is the gravitational potential at the location of the host star in the cluster. The value of $\phi_*$ at the time of escape is obtained from \nbo{}. Equation~(\ref{eq:discharging}) can then be written as 
\begin{equation}
    v_\mathrm{discharge} = \sqrt{-2\phi_\mathrm{particle}} \approx \sqrt{-2\phi_*}
    ~\quad .
    \label{eqn.vdischarge}
\end{equation}
This is the minimum speed required for a particle to be discharged from the star cluster. Below, we will refer to the {\em local} $v_\mathrm{discharge}$, to emphasize that it is a local property of the star cluster, and that is also depends on time. Note that the local $v_\mathrm{discharge}$ differs for each escape event.

We compare the escape speed of escaped \pdps{} ($v_\mathrm{escape,sc}$) with the local discharge barrier speed of the star cluster (local $v_\mathrm{discharge}$), to determine whether the particle escaped from the planetary system will be discharged from the star cluster. 
Figure~\ref{fig.8k_CDF_escapers_speed_over_vesc} shows the corresponding cumulative distribution function of $v_\mathrm{escape,sc}~/~\mathrm{local}~v_\mathrm{discharge}$. Among the escaped \pdps{}, 66.3\% will remain gravitationally bound to the star cluster as free-floating debris particles. These particles will eventually be discharged from the star cluster, while some may be dynamically expelled from the star cluster, or may even be re-captured by other stars. 33.7\% of the escaped \pdps{} will be discharged immediately from the star cluster. The presence of the planet has no noticeable influence on this fraction, as the population of escapers is dominated by \pdps{} from the outer regions of the planetary system. Note, however, that this fraction depends on the initial semi-major axis range of \pdps{}, which determines the probability of escape from the planetary system. Among the 16 planets that escape from their host stars, 14 will immediately leave the cluster, while 2 remain bound to the star cluster. The dynamics of a planet (in a single-planet system) is very similar to that of a \pdp{} with an identical $a_0$ (in a system without planets), as the masses of both types of particles are much smaller than those of stars.

We proceed with studying the escape speed $v_\mathrm{escape,sc}$ and also the value $v_\mathrm{escape,sc}~/~\mathrm{local}~v_\mathrm{discharge}$ as a function of the initial semi-major axis, as shown in Figure~\ref{fig.8k_0_escapers_speed_over_a0all}. It is worth noticing that the number of escapers from the inner orbits is generally much smaller than those from the outer orbits (except for orbits near the planet), so the statistical significance of the data points on the right-hand side is larger. Additionally, the number of escapers is initially small, so the statistical confidence of the curves increases with time.

In the absence of the planet, the fraction of discharged \pdps{} tends to decrease with increasing initial semi-major axis. The escapers in the regions close to the host star all have higher escape speeds, although the number of escapers is small (see Figure~\ref{fig.8k_esc_fraction_over_a0_by_t}). For particles with $a_0<75$~au, most ($>75\%$) of the escapers are immediately discharged from the cluster, while more than half of the escapers with $a_0<300$~au gain enough speed to be discharged. A transition occurs in the region between 300~au and 350~au, with about half of the particles remaining bound to the cluster and the other half escaping.

An interesting result is that, in models where the \fifju{} is present, the escape speed of interior particles ($a_0 \le 150$~au) is lower, causing more to remain in the cluster. This is notably the case for the region \textit{private region of the planet} ($40-60$~au), more than half of the particles remain in the cluster.  In Section~\ref{sec.esc_fracion_over_a0}, we have found that the planet kicks \pdps{} from the planetary system (Figure~\ref{fig.8k_esc_fraction_over_a0_by_t}), but these kicks are \textit{gentle}, so that particles are pushed outside the planetary system but still remain members of the star cluster. The planet has a small influence on the escape speeds of exterior particles, and no effect on particles outside \textit{reach of the planet} ($a_0 > 400$~au). The findings discussed above are valid for $t=100$~Myr, but they also apply to $1<t<100$~Myr, when $a_0 \geq 100$~au. Statistics of escapers are insufficient for other values of $t$ and $a_0$.

The host stars are at different locations in the star cluster when the individual \pdps{} escape. However, the curves in the upper and lower panels of Figure~\ref{fig.8k_0_escapers_speed_over_a0all} are similar. This indicates that $v_\mathrm{discharge}$ is roughly similar for \pdps{} at their time of escape. This suggests that most escapers come from similar distances to the cluster center.

We do not track particles that have escaped from the planetary systems. Although we identify particles with $v_\mathrm{escape,sc} > \mathrm{local}~v_\mathrm{discharge}$ as discharged from the cluster, \pdps{}' interactions with the local stellar environment may result in a different evolution. Scattering events with neighboring stars may slow down or speed up the escaped \pdps{}, and re-capture by another star may occur. Capture and re-capture may be rare according to \citet{parker2012, perets2012}, but a recent study by \citet{hands2019} argues that re-capture of free-floating planetesimals in the star cluster is common.

\citet{cai2019} studied planetary systems with debris disks (massless particles from 6~au to 400~au orbiting solar mass stars) in young massive clusters ($N=128$k, $\hmrin{}=1.34$pc). Their figure~9 shows $v_\mathrm{escape,sc}~/~\mathrm{local}~v_\mathrm{discharge}$ versus of ejection semi-major axis, colored by the initial semi-major axis. Considering that their $a_0$ and $a_\mathrm{eject}$ correlate positively, their result also shows the tendency that (i) escapers from the inner region generally move fast, and are more likely to be discharged from the star cluster (ii) escapers from the outer region are likely to remain part of the star cluster. \citet{cai2019} argue that the \textit{prompt ejectors}, which in our words are the particles discharged from the star cluster, originate primarily from regions with small $a_0$. Although in our study we also observe the correlation between discharge probability and initial semi-major axis, the problem may lie in the quantity: in Section~\ref{section:esc_fraction} (e.g., Figure~\ref{fig.8k_esc_fraction_over_a0_by_t}) we find that the number of escapers rises with increasing initial semi-major axis. Moreover, the results also depend on the properties of the stellar environment (i.e., the star cluster). Further studies are required to constrain the origin and evolution of free-floating particles in star clusters, and how star clusters can provide a source of interstellar objects, such as 1I/’Oumuamua \citep{oumuamua}, and the recently confirmed CNEOS 2014-01-08 \citep{CNEOS2014}.

%% file: sections/conclusion.tex
\section{Discussion and Conclusions}
\label{section:conclusions}

Most stars and their planetary systems are formed in dense stellar environments, where close encounters with neighboring stars are common. In this study, we present an investigation into how the presence or absence of a planet affects the long-term evolution of planetary debris particles in planetary systems that are embedded in star clusters. We carry out $N$-body simulations using \nbo{} and \reb{}, using an optimization of the \lps{} scheme. Our main findings for the evolution of planetary systems around solar-mass stars in our modeled star cluster ($N=8000$, $\hmrin{}=0.78$~pc) are summarized as follows.

\begin{enumerate}[leftmargin=0.6cm, labelsep=3pt, itemindent=-5pt]
    
    \item Based on the outcomes of the simulations, we identify three regions in the planetary system (see Figure \ref{fig.division}): (1) \textit{the private region of the planet} $40-60$~au, where no \pdp{} survive because of the planet clearing out its nearby orbits; (2) \textit{the reach of the planet} $0-400$~au, an area where particles are influenced by the \fifju{} and (3) \textit{the territory of the planetary system}, most particles outside of which will eventually escape into the star cluster. This classification allows a rough prediction of the planet's influence on the escape fractions, the escape speeds, and the evolution of the orbital elements of the \pdps{}. 
    
    \item The presence of the \fifju{} increases the escape fraction of the \pdps{}. The magnitude of this trend depends on the \pdp{}'s initial semi-major axis, $a_0$. The influence of the planet is most pronounced in \textit{the private region of the planet} ($40 \le a_0/\mathrm{au} \le 60$), within which most \pdps{} are quickly removed, where more than 85\% are removed by 25~Myr and more than 90\% by 100~Myr. Another locality that is strongly affected is the 2:1 mean motion resonance location at 80~au, which yields a 77.0\% escape fraction at 100~Myr when the planet is present, more than three times the value of 25.1\% when the planet is absent. \Pdps{} at other $a_0$ are affected to a lesser extent, with a 50\% increase in the escape fraction for $a_0 < 100$~au and less than 50\% for $100-400$~au, compared with simulations without the planet. The escape fraction of \pdps{} as a function of time can be described with a hyperbolic function (Equation~\ref{eqn.fit}), which predicts that, given sufficient time, all particles with $a_0 \ge 700$~au will eventually escape from the planetary system. The difference in escape fraction between systems with and without a planet, $\Delta\chi$, grows roughly linearly with time, before it reaches a limit that depends on $a_0$. 

    \item The planet induces changes in the orbital elements of the surviving \pdps{}. The degree of change depends on the initial semi-major axis. In simulations without a planet, interactions with neighboring stars raise the eccentricities and inclinations of the \pdps{} in the interior region, while the corresponding semi-major axis distribution changes slightly. When the \fifju{} is present, a significant increase in the fraction of particles with high eccentricity and inclination is seen, and nearly all \pdps{} around 50~au are removed. Most \pdps{} obtain $e=0-0.1$ when a planet is present, with a nearly uniform distribution in this range, which is also seen in isolated planetary system simulations. For the exterior region, few particles remain in orbits with large semi-major axes, and the eccentricity distribution is nearly uniform in the range $e \in [0,1]$, and compared to the interior region, more particles obtain high inclinations. In both the interior and exterior regions, the presence of the planet decreases the fraction of particles on retrograde orbits and on inclinations above the Kozai angle. The orbital distributions of \pdps{} in the exterior region are independent of whether the planet is present, although the planet appears to accelerate the evolution in both the interior and exterior regions. When the planet is present, more surviving particles experience outward migration, and fewer experience inward migration. 

    \item Most of the high-velocity escapers from the planetary system originate from the hot region of the planetary system, and some of these have high enough speeds that allows direct escape from the star cluster. Low-velocity escapers originate mostly from the cold region of the planetary system. The presence of the planet reduces the escape speed from the interior region, so that a higher fraction among these remain bound to the star cluster. Overall, $\sim 66\%$ of the particles remain bound to the star cluster. These become free-floating particles that may get re-captured by other stars or may eventually escape from the star cluster. 

\end{enumerate}

In this work, we simulate both the star cluster and planetary systems for 100~Myr. This timescale is adequate for studying (i) the evolution of the escape fraction for most particles (Figure~\ref{fig.8k_esc_fraction_over_a0_by_t}), (ii) the influence of the planet on the \pdps{} (Figure~\ref{fig.diff_Nesc_over_a0_by_t_8k}) except those initially on $60-150$~au, (iii) the evolution of the orbital element distributions of the surviving \pdps{}  (Figure~\ref{fig.8k_SURVIVORS}), and (iv) the escape speeds (Figure~\ref{fig.8k_0_escapers_speed_over_a0all}). A timespan of 100~Myr is thus appropriate for characterizing the main features of the dynamical evolution of the planetary systems studied in this work.

We have made several approximations and simplifications that may warrant further study. In this work, we have not included primordial stellar binaries. Stellar binaries have substantially larger collisional cross-sections than single stars \citep[e.g.,][]{liadams2015}, and may therefore increase the escape fraction of \pdps{} and planets \citep[see, e.g.,][]{wangyh2020}. We have only focused on solar-mass host stars. Planetary systems around more massive stars can be vulnerable to disruption as a consequence of mass segregation (massive stars tend to sink into the cluster center, where the stellar encounter rate is higher), stronger gravitational focusing, and stellar evolution. \citet{hands2019} studies debris disks in small open clusters, and found massive stars are more likely to lose planetesimals than their lower-mass siblings. \citet{stock2022} studied planetary system around host stars of different masses in star clusters, and found the fraction of perturbed planets increases for host star mass of $2.5~\mdotin{}$ compared with $1.5~\mdotin{}$ host stars. 

In terms of planetary system architecture, we have only studied systems containing one Jupiter-mass planet at 50~au. Previous studies on isolated planetary systems \citep{nesvold2015, tabeshian2016, tabeshian2017} have constrained the debris disk's response to different planetary masses, eccentricities, and inclinations. Based on this, we speculate that (i) the width of the peak in the escape fraction near 80~au (2:1 resonance) is positively correlated with planet's mass (while an over-massive planet will eliminate the peak), and is anti-correlated with planet's eccentricity; (ii) the escape fraction peak near 104~au (3:1) will be more prominent when the planet's eccentricity is higher; and (iii) slightly increasing the inclination of the planet (of the order of Jupiter's inclination) makes little difference, while a highly-inclined planet (e.g., $i=30^\circ$) may be able to disrupt the entire debris disk. Moreover, the effects of the planet and the stellar environment on the debris disk are not independent. The evolution of debris disks in planetary systems with multiple planets, with a single planet in a different orbital configuration, and in different star cluster environments, will be addressed in our future studies.